\documentclass{elsart}
\usepackage{amsfonts}
\usepackage{amsmath}
\usepackage{amssymb}
\usepackage{graphicx}
\usepackage{epsfig}
\begin{document}
\begin{frontmatter}

\title{The Field Theory Approach to Percolation Processes}

\author{Hans-Karl Janssen}
\address{Institut f\"ur Theoretische Physik III, 
     Heinrich-Heine-Universit\"at, \\ 40225 D\"usseldorf, Germany}

\author{Uwe C. T\"auber}
\address{Department of Physics, 
     Virginia Polytechnic Institute and State University, \\ 
         Blacksburg, VA 24061-0435, USA}

\date{\today}

\begin{abstract}
We review the field theory approach to percolation processes. Specifically, we
focus on the so-called simple and general epidemic processes that display
continuous non-equilibrium active to absorbing state phase transitions whose
asymptotic features are governed respectively by the directed (DP) and dynamic
isotropic percolation (dIP) universality classes. We discuss the construction 
of a field theory representation for these Markovian stochastic processes based
on fundamental phenomenological considerations, as well as from a specific 
microscopic reaction-diffusion model realization. Subsequently we explain how 
dynamic renormalization group (RG) methods can be applied to obtain the 
universal properties near the critical point in an expansion about the upper
critical dimensions $d_c = 4$ (DP) and $6$ (dIP). We provide a detailed 
overview of results for critical exponents, scaling functions, crossover 
phenomena, finite-size scaling, and also briefly comment on the influence of 
long-range spreading, the presence of a boundary, multispecies generalizations,
coupling of the order parameter to other conserved modes, and quenched 
disorder.
\end{abstract}

\end{frontmatter}

\section{Introduction}

\subsection{Percolation Processes}

\begin{figure}[b]
\begin{center}\includegraphics[width=6cm]{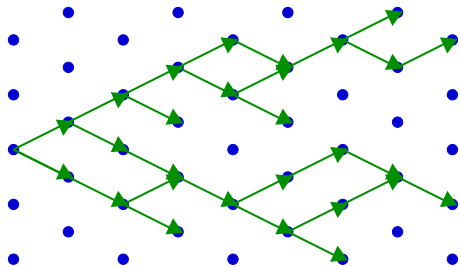}\end{center}
\caption{Directed percolation (DP) in two dimensions: bonds connecting the
         lattice sites can only be passed from left to right.}
\label{dirper}
\end{figure}

The investigation of the formation and the stationary properties of random
structures has been an exciting topic in statistical physics for many years.
Since it provides an intuitively appealing and transparent model of the
irregular geometry emerging in disordered systems, \emph{percolation} has
provided a leading paradigm for random structures. \emph{Bond percolation}
constitutes perhaps the simplest percolation problem. Two fundamental variants
of bond percolation have been introduced in the past: In the \emph{isotropic
percolation} (IP) problem the bonds connecting the sites of a regular lattice
(in $d$ spatial dimensions) are randomly assigned to be open (with probability
$p$) or blocked (with probability $1-p$), and an agent may traverse an open
bond in either direction. In contrast, in the \emph{directed percolation} (DP)
problem the open bonds can be passed only from one of the two connecting
sites, whence the allowed passage direction globally defines a preferred
direction in space, see FIG.~\ref{dirper}.

\begin{figure}[ptb]
\begin{center}{\epsfxsize = 6 truecm \epsffile{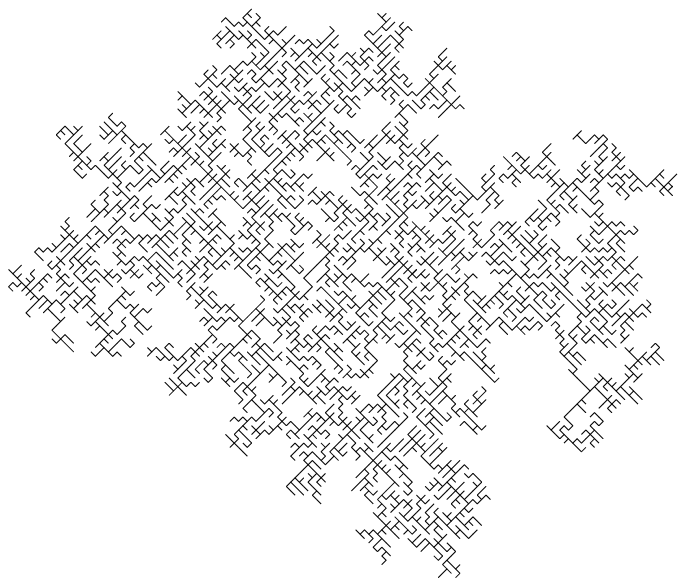}
               \epsfxsize = 2.5 truecm \epsffile{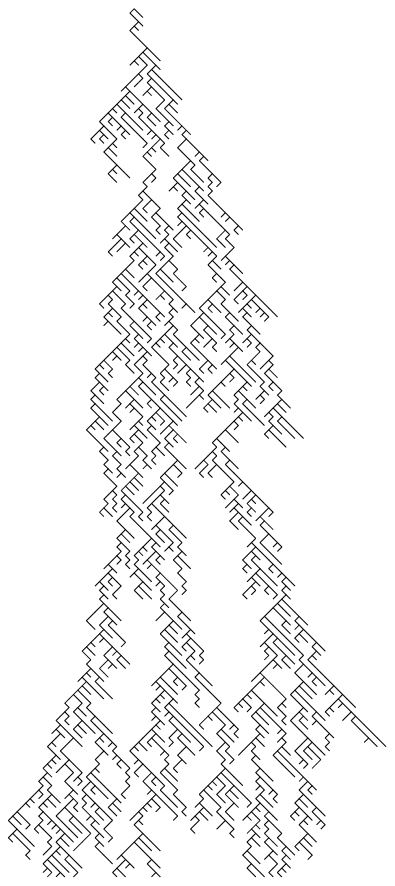}}\end{center}
\caption{Two-dimensional critical isotropic (left) and directed (right, with 
         preferred direction down) percolation clusters. 
     (From Ref.~\cite{FTS94}).}
\label{perccl}
\end{figure}

In these terms, percolation can be viewed as the passage of an agent through
an irregularly structured medium, in the sense that the agent can propagate
through certain regions, whereas it cannot traverse other areas. Though
percolation represents one of the simplest models for random systems, it has
in fact many applications. Moreover, it yields a prototypical
\emph{non-equilibrium phase transition}: For small values of $p$, regions with
open bonds form disconnected clusters of typical linear extension $\xi$,
wherein the agent becomes localized or trapped. For $p > p_c$, a critical
threshold value for the frequency of open bonds, the probability for the
existence of an infinite connected accessible cluster becomes non-zero.
Therefore, an agent on an infinite cluster is not confined to a finite region
of space any more, but may percolate through the entire system. This
percolation transition defines a genuine \emph{critical phenomenon} as
encountered in equilibrium statistical mechanics, with $\xi$ providing the
divergent length scale as $p_c$ is approached. It turns out that many,
microscopically quite different percolation-type systems share their critical
properties either with IP or DP. Thus, the phase transitions in IP and DP, the
latter distinguished from the former through the broken spatial isotropy,
define genuine \emph{universality classes}. For reviews on isotropic and
directed percolation the reader is referred to
Refs.~\cite{StAh92BuHa96,Ki83,Hin00}. Critical IP and DP clusters are shown
in FIG.~\ref{perccl}.

Broadbent and Hammersley \cite{BrHa57} first directed attention to the
\emph{dynamics} of percolation processes. In contrast to random diffusion, in
percolation processes the spreading of agents is mainly governed by the
disordered structure of the medium. Consequently, they define stochastic
growth processes where agents randomly generate offspring at neighboring
sites, with rates determined by the susceptibility of the medium. The agents
themselves decay spontaneously, thereby producing debris. In this way the
medium becomes exhausted, and its susceptibility is randomly diminished.
Hence, static DP in $d+1$ spatial dimensions can be directly interpreted as a
Markov process in $d$ spatial and $1$ time dimension. The preferred
(propagation) direction defines the temporal axis, and the history of the
process yields the debris that forms the ($d+1$)-dimensional spatially
anisotropic \emph{directed} percolation clusters. In contrast, in a
\emph{dynamical isotropic percolation} (dIP) process the agents grow from a
seed in an expanding stochastic annulus in $d$ spatial dimensions leaving
behind the debris in their wake. This debris then forms $d$-dimensional
\emph{isotropic} percolation clusters.

Remarkably, the renormalized field theory of DP has appeared originally in
elementary particle physics in the guise of \emph{Reggeon field 
theory}~\cite{Gri67,GriMi68,Mo78}. Grassberger et al.~\cite{GraSu78,GraTo79} 
pointed out that Reggeon field theory is not a Hamiltonian field theory but
constitutes a stochastic process for which they coined the name \emph{Gribov
process}. It represents a stochastic version of Schl\"ogl's `first reaction'
\cite{Schl72}. Subsequently, the formal connection of Reggeon field theory to
DP was explicitly demonstrated \cite{Ob80,CaSu80,Ja81}, and Janssen and
Grassberger stated the \emph{DP conjecture}: The critical behavior of an order
parameter field with Markovian stochastic dynamics, decoupled from any other
slow variable, that describes a transition from an active to an inactive,
\emph{absorbing state} (where all dynamics ceases) should be in the DP
universality class \cite{Ja81,Gr82}. The field theory of dIP was initiated by 
Grassberger's formulation of the \emph{general epidemic process} (GEP) on a 
lattice, and his statement that near criticality this process produces
isotropic percolation clusters \cite{Gr83}. This insight led to the 
construction of the renormalized field theory for the dIP universality class 
\cite{Ja85,CaGra85}.

\subsection{Universality Principles of Percolation}

\begin{figure}[ptb]
\begin{center}\includegraphics[width=6.5cm]{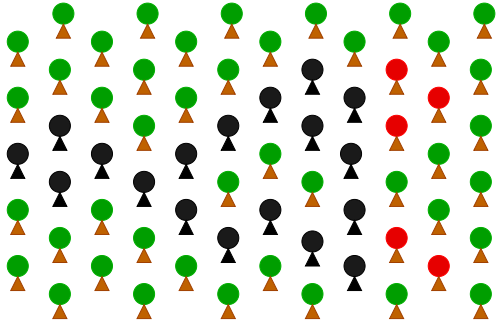}\end{center}
\caption{Directed percolation represented as a forest fire under the influence 
         of a storm from left to right; burning trees are depicted as red, and 
     burned dead trunks are shown black.}
\label{waldbrand}
\end{figure}

It is tempting to express percolation processes in the language of an epidemic
disease such as blight in a large orchard, the spreading of bark beetles in a
forest, or proliferation of a forest fire, as depicted in FIG.~\ref{waldbrand}.
Here, the susceptible individuals or healthy trees form the medium, and the 
disease, blight, beetles, or fire represent the agent. The sick individuals, 
the befallen or burning trees may randomly infect neighboring individuals. The 
sick individuals are allowed to recover, becoming susceptible anew. This 
behavior defines the so-called \emph{simple epidemic process} (SEP), also known
as \emph{epidemic with recovery} \cite{Mur89}. Properties of the SEP near the 
ensuing critical point separating the endemic and pandemic phases (whose 
precise location depends on the susceptibility of the individuals) shares the 
universal properties of DP. In contrast, in a situation where the sick 
individuals die out or become immune as opposed to susceptible again, the 
medium is eventually exhausted. This scenario defines the \emph{general 
epidemic process} (GEP), also termed \emph{epidemic with removal} 
\cite{Mur89,Mol77,Bai75}. Its universal properties are governed by the dIP 
universality class. The statistical properties of the debris clusters that are
left behind after the disease is extinguished are described by static IP.

In the recent physical literature, basically two different methods have been
employed to construct field theories of percolation processes. The first
approach starts from a specific microscopic stochastic reaction-diffusion
model formulated in terms of a master equation on a lattice. Basically it
consists of a description of the random walks generated by diffusing and
reacting particles. A field theory is then obtained through a representation of
the master equation by means of a bosonic creation/destruction operator 
formalism and introduction of coherent-state path integrals, followed by a 
naive continuum limit \cite{Doi76,GrSch80,Pel84,Lee94,HoTaVo04}. The second 
method, to be applied in the first part of this article, takes a more 
universal, phenomenological perspective, and can be viewed as a dynamical 
analog of the established Landau-Ginzburg-Wilson approach to static critical 
phenomena. It is based on formulating the fundamental principles shared by all 
systems in a given universality class, and directly employs a \emph{mesoscopic}
stochastic continuum theory. Of course, one has to a-priori define the slow 
dynamic fields in the theory, typically local densities of the order parameter 
and any conserved quantities, as characterizing the universality class under
consideration. This approach assumes that on a mesoscopic scale the motion of
the fast microscopic variables can be absorbed into the statistical properties
of the stochastic equations of motion for the relevant slow variables. The
corresponding stochastic processes should be Markovian provided all slow
variables are retained. Taking into account the general (symmetry, 
conservation, etc.) properties that define the universality class, the
stochastic field theory is constructed using small-density and gradient
expansions, i.e., long-wavelength and low-frequency approximations that focus
on the infrared (IR) properties of the system.

For percolation processes, the fundamental statement regarding the slow
variables consists in the assertion that percolation near the critical point
can be described by a Markov process that only depends upon the density
$n(\mathbf{r},t)$ of activated sick parts of the medium and, in the case of
dIP, upon the density of the debris $m(\mathbf{r},t)$. In terms of the field
$n$ the density of the debris is given by $m(\mathbf{r},t) = \lambda
\int_{-\infty}^{t} n(\mathbf{r},t^{\prime}) \, dt^{\prime}$, where $\lambda$
denotes an appropriate kinetic coefficient. Indeed, this assumption that
percolation can be described solely in terms of $n$ becomes manifest in
typical lattice simulations of such processes. Near the critical point one
observes a patchwork of regions consisting of \emph{correlated} active
neighboring sites and vacuum regions devoid of any activity. In stark contrast
to this picture, simulations for branching and annihilating random walks show
anticorrelating behavior, namely widely separated single active sites
propagating on isolated paths and reacting only after encounter
\cite{Hin00,Od04}.

We now proceed to formulate four principles that allow the explicit
construction of a mesoscopic stochastic description of percolation processes.
We shall describe the universal aspects of the dynamics of $n$ beginning with
the DP universality class:

\begin{enumerate}
\item[(i)] The susceptible medium becomes locally infected, depending on the
density $n$ of neighboring sick individuals. The infected regions recover
after a brief time interval.

\item[(ii)] The state with $n \equiv0$ is absorbing. This state is equivalent
to the extinction of the disease.

\item[(iii)] The disease spreads out diffusively via the short-range infection
(i) of neighboring susceptible regions.

\item[(iv)] Microscopic fast degrees of freedom may be captured as local noise
or stochastic forces that respect statement (ii), i.e., the noise alone cannot
regenerate the disease.
\end{enumerate}

For the dIP universality class, we need to modify principles (i) and (ii) by

\begin{enumerate}
\item[(i')] The susceptible medium becomes infected, depending on the
densities $n$ \emph{and} $m$ of sick individuals and the debris, respectively.
After a brief time interval, the sick individuals decay into the immune
debris. The debris ultimately stops the disease locally by exhausting the
supply of susceptible regions.

\item[(ii')] The states with $n \equiv0$ and any spatial distribution of $m$
are absorbing. These states are equivalent to the extinction of the disease.
\end{enumerate}

It is important to realize that the mechanism (i') introduces \emph{memory}
into the stochastic process. Note that we do not explicitly assume diffusive
motion of the sick individuals. All that is needed is a spreading mechanism
for the agent, here the disease itself: $\partial_{t} \, n(t,\mathbf{r}
)|_{\mathrm{spread}} = \int d^{d}r \, P(\mathbf{r}-\mathbf{r}^{\prime}) \,
n(t,\mathbf{r}^{\prime})$ with a spreading probability $P(\mathbf{r})$ that is
assumed short-range, hence allowing to approximate its propagation by simple
diffusion. However, this scenario by no means excludes reaction-diffusion
systems where the ``infected'' particles themselves spread out diffusively, if
otherwise the general percolation principles are satisfied.

\section{Field Theory Representation of Percolation Processes}

\subsection{Dynamic Response Functional for Stochastic Percolation Processes}

We proceed to set up the stochastic equations of motion for percolation 
processes according to the general principles formulated in the preceding 
section,
\begin{equation}
\label{EqMot}
  \partial_t \, n(\mathbf{r},t) = \mathcal{V}(\mathbf{r},t) \, , \quad 
  \partial_t \, m(\mathbf{r},t) = \lambda \, n(\mathbf{r},t) \, ,
\end{equation}
or, in a discretized version (in the It\^o sense) $\{t\} \to \{t_k\}$ 
with $t_{k+1} = t_k + \Delta$,
\begin{equation}
\label{DiscEM}
  n(\mathbf{r},t_{k+1}) = n(\mathbf{r},t_k) 
  + \Delta \mathcal{V}(\mathbf{r},t_k) \, , \quad
  m(\mathbf{r},t_{k+1}) = m(\mathbf{r},t_k) 
  + \lambda \, \Delta n(\mathbf{r},t_k) \, .
\end{equation}
Here, $\mathcal{V}(\mathbf{r,}t)$ denotes a random density that has yet to be
constructed, and the second equations for $m$ only appear for the dIP process.
For simplicity we shall suppress spatial arguments in the following
considerations. The Markovian assumption implies that the stochastic variables 
$\mathcal{V}(t_k)$ are uncorrelated at different times, and that the 
statistical properties of $\mathcal{V}(t_k)$ depend solely upon $n(t_k)$ and 
$m(t_k)$. As a consequence, the generating function of the cumulants of 
$\mathcal{V}$ (i.e., the Laplace transform of the corresponding probability
distribution) must have the general form
\begin{equation}
\label{Kum-V}
  \overline{\exp \Big( \Delta \sum_k \mathcal{V}(t_k) \, \tilde{n}(t_{k+1}) 
  \Big)} = \exp \Big( \Delta \sum_k \sum_{l_k=1}^\infty 
  \frac{\tilde{n}(t_{k+1})^{l_k}}{l_k !} \, K_{l_k}[n(t_k),m(t_k)] \Big) \, ,
\end{equation}
where the overbar denotes the statistical average over the fast microscopic
degrees of freedom. The $\tilde{n}(t_k)$ constitute independent new variables. 
Of course, in the DP process the cumulants $K_l$ are independent of the debris 
$m$.

The statistical properties of the stochastic process are fully encoded in the 
simultaneous probability density of the history
$\{ n(t_0),n(t_1),\ldots,n(t_k),\ldots \}$:
\begin{align}
\label{SimProbal}
  \mathcal{P}(\{n(t)\}) &= \overline{\prod_k 
  \delta\bigl[ n(t_{k+1})-n(t_k) - \Delta\mathcal{V}(t_k) \bigr]} \\
  & = \int \prod_k \frac{d\tilde{n}(t_{k+1})}{2\pi i} \ \overline{\exp \Big\{
  \sum_k \tilde{n}(t_{k+1}) \bigl[ \Delta \mathcal{V}(t_k)+n(t_k)-n(t_{k+1})
  \bigr] \Big\}} \, . \nonumber
\end{align}
Upon inserting (\ref{Kum-V}) and reverting to continuous time, we may then 
formally write $\mathcal{P}$ as a path integral over functions $\tilde{n}(t)$:
\begin{equation}
\label{SimProb}
  \mathcal{P}(\{n(t)\}) = \int \mathcal{D}[\tilde{n}] \, \exp \int \! dt \,
  \biggl\{ \, \sum_{l=1}^\infty \frac{\tilde{n}(t)^l}{l !} \, K_l[n(t),m(t)] 
  - \tilde{n}(t) \, \partial_t n(t) \, \biggr\} \, .
\end{equation}
This expression is always to be interpreted through the preceding prepoint 
(It\^o) discretization; we note that this specifically implies 
$\theta(t \leq 0) = 0$ for the Heaviside function that enters the dynamic 
response functions. The integration over the `response variable' $\tilde{n}$ 
runs along the imaginary axis from $-i\infty$ to $+i\infty$. However, this path can be analytically deformed in finite regions. Therefore $\tilde{n}$ can attain real contributions.

All cumulants $K_l$ are subject to a fundamental constraint as a consequence of
the existence of the absorbing state(s): they must vanish for $n = 0$. Assuming
that the functionals $K_l$ can be expanded in powers of $n$ and $m$, this 
implies that all $K_l(n,m) \propto n$ or higher powers of $n$. Moreover, as we
shall demonstrate below, the higher cumulants $K_l$ with $l \geq 3$ are 
irrelevant in the RG sense. Thus, the remaining task is to find the functional
forms of the mean-field part $K_1$ and the Gaussian stochastic force correlator
$K_2$. After reducing the statistical properties of $\mathcal{V}$ to these two 
terms, the stochastic equation of motion for $n$ may be written in the usual 
Langevin form (in the It\^o interpretation)
\begin{subequations}
\label{LangevGl}
\begin{align}
  \partial_t \, n(t) & = K_1[n(t),m(t)] + \zeta(t) \, , \label{LangevGl1} \\
  \overline{\zeta(t) \, \zeta(t^\prime)} & = K_2[n(t),m(t)] \, 
  \delta(t-t^\prime) \, . \label{LangevGl2}
\end{align}
\end{subequations}
Note that $K_2 \geq 0$ since our theory is based on a real field $n$.
This is in contrast to the path integral representation of the theory of random
walks subject to pair annihilation which may formally lead to a negative second
cumulant \cite{Lee94,HoTaVo04}.

Statistical averages of any functional of the slow variable $n$ can now be
performed by means of path integrals over the fields $n$ and $\tilde{n}$ with 
the weight $\exp(-\mathcal{J})$ \cite{Ja76,DeDo76,Ja92}; reinstating the 
spatial arguments, the integral measure here is $\mathcal{D}[\tilde{n},n] = 
\prod_{\mathbf{r},t} \, d\tilde{n}(\mathbf{r},t) \, dn(\mathbf{r},t) / 2\pi i$.
Retaining only $K_1$ and $K_2$, the \emph{dynamic response functional} 
$\mathcal{J}$ from Eq.~(\ref{SimProb}) becomes
\begin{equation}
\label{J1}
  \mathcal{J} = \int dt \, \Big\{ \tilde{n}(t) \bigl[ \partial_t \, n(t) 
  - K_1[n(t),m(t)] \bigr] - \frac{1}{2} \, \tilde{n}(t)^2 \, K_2[n(t),m(t)] \,
  \Big\} \, .
\end{equation}
Here integrations over $d$-dimensional space are implicit.

Fundamental principles (ii) and (ii') classify DP and dIP as absorbing state
systems: Any finite realization will reach an absorbing state in a finite time
$\tau_{\rm abs}$ with probability $1$. However, in the active pandemic state 
above the critical percolation threshold $p_c$, $\tau_{\rm abs}$ grows 
exponentially with the system size $L^d$. It is thus appropriate to set 
$\tau_{\rm abs} \to \infty$ in the thermodynamic limit $L \to \infty$. The
assertion that principles (i)--(iv) lead to the DP universality class is known 
in the literature as the DP conjecture \cite{Ja81,Gr82}. The diffusive 
spreading of the disease renders the cumulants $K_1$ and $K_2$ \emph{local} 
functionals of $n$ and $m$, whence a gradient expansion is appropriate. The 
absorbing state condition then inevitably implies
\begin{equation}
 \label{K1}
  K_1(n,m) = R(n,m) \, n + \lambda \, \nabla^2 n + \ldots \, ,
\end{equation}
where the ellipsis denotes higher-order gradient terms. In the same manner,
\begin{equation}
\label{K2}
  K_2(n,m;\mathbf{r}-\mathbf{r}^\prime) = 2 \bigl[ \Gamma(n,m) \, n 
  + \lambda^\prime \, (\nabla^2 n) - \lambda^{\prime\prime} \, n \nabla^2 
  + \ldots \bigr] \, \delta(\mathbf{r}-\mathbf{r}^\prime) \, .
\end{equation}
The contributions with kinetic coefficients $\lambda^\prime$ and $
\lambda^{\prime\prime}$ can be interpreted as purely diffusional noise if
$\lambda^{\prime\prime} = 2 \lambda^\prime$. Yet here these terms simply arise 
from the gradient expansion. We shall demonstrate below that these as well as 
higher-order terms are in fact irrelevant for the critical properties of 
percolation. Naturally, the dependence of $m$ is absent for DP. 

Furthermore, $n$ and $m$ are small quantities near the critical point. This 
allows for a low-density expansion 
\begin{equation}
\label{dens_exp}
  R(n,m)  = - \lambda \, \bigl( \tau + g_1 \, n + g_2 \, m + \ldots \bigr) \, ,
  \quad \Gamma(n,m) = \lambda \, \bigl( g_3 + \ldots \bigr) \, ,
\end{equation}
with $g_2 = 0$ for DP. To ensure stability, the coupling constants $g_i$ are
assumed to be positive; otherwise one has to extend the expansion to the 
subsequent powers. Note that there is no general reason for $g_3$ to vanish as
is the case for interacting random walks \cite{Lee94,HoTaVo04}. Discarding the 
irrelevant higher-order expansion terms, indicated by the ellipsis in 
Eqs.~(\ref{K1})--(\ref{dens_exp}), and inserting into Eq.~(\ref{J1}), we at 
last arrive at the fundamental dynamic response functional for percolation 
processes:
\begin{equation}
\label{J2}
  \mathcal{J} = \int d^dr \, dt \, \Big\{ \tilde{n} \bigl[ \partial_t 
  + \lambda \bigl( \tau-\nabla^2 \bigr) + \lambda \bigl( g_1 \, n + g_2 \, m 
  - g_3 \, \tilde{n} \bigr) \bigr] n - q \, \tilde{n} \Big\} \, .
\end{equation}
Here we have introduced an additional external source $q = q(\mathbf{r},t)$ for
the agent. Specifically, a seed inserted at the origin $\mathbf{r} = 0$ at time
$t = 0$ in order to initialize a spreading process is modeled by
$q(\mathbf{r},t) = \delta(\mathbf{r}) \, \delta(t)$. The evaluation of the path
integrals is always restricted by the initial and final conditions 
$n(t \to -\infty) = 0 = \tilde{n}(t \to +\infty)$, respectively. After 
eliminating the fast microscopic degrees of freedom, and thereby focusing on 
the IR properties of the relevant slow variables, the wave vectors $\mathbf{q}$
of the fluctuations of the Fourier transformed fields $n_\mathbf{q}(t)$ and
$\tilde{n}_\mathbf{q}(t)$ is clearly limited by a momentum cutoff $\Lambda$. 
Correspondingly, those fluctuations are effectively set to zero for 
$|\mathbf{q}| \gg \Lambda$.

\subsection{Master Equation Field Theory for a  Specific Reaction-Diffusion 
            Model}

An alternative approach to constructing a field theory representation for a
stochastic system starts from the classical master equation that defines the
process microscopically. For reaction-diffusion systems, one may formulate the
reactions in a straightforward manner in terms of (bosonic) creation and
annihilation operators, and therefrom via coherent-state path integrals proceed
to a field theory action \cite{Doi76,GrSch80,HoTaVo04}. This method invokes
no phenomenological assumptions or explicit coarse-graining, but does rely on a
the validity of a `naive' continuum limit. One may thus view this procedure in
analogy to the derivation of the $\Phi^4$ theory from a soft-spin Ising model 
on a lattice by directly taking the limit of zero lattice constant. Universal
features do not immediately become apparent in such a treatment, whereas the 
approach laid out in the preceding section focuses on general principles and in
this sense rather corresponds to the Landau-Ginzburg-Wilson model for 
second-order equilibrium phase transitions. The phenomenological construction
certainly requires considerable a-priori insight on the defining universal
features of a given system, but its benefit is considerable predictive power.
Yet there are cases, such as pure pair annihilation processes 
\cite{Pel84,Lee94} or branching and annihilating random walks with even 
offspring \cite{CaTa96}, for which as yet only the master equation formulation 
has led to successful field theory representations. In these processes the 
random-walk properties of individual particles govern the essential physics;
hence the microscopic picture is essential and a mesoscopic description simply
in terms of the particle density would be erroneous. On the other hand, 
proceeding from the microscopic model via taking a naive continuum limit 
definitely fails, e.g., for the pair contact process with diffusion (PCPD)
\cite{JavWDeTa04,HeHi04}.

We shall consider here a specific reaction-diffusion model that displays an
active to absorbing state phase transition which, according to our general 
principles, should belong to the percolation universality classes. Yet 
reaction-diffusion systems do of course not provide an exact description of, 
e.g., directed bond percolation or the contact process since they differ
manifestly with respect to the microscopic details. We note that one may also 
construct a direct field theory representation for the pair connectivity of 
bond-percolating system \cite{CaSu80,BeCa84}. As usual in reaction-diffusion 
models, the activated particles both represent the diffusing agents and may
locally create new offspring. We introduce the following reaction scheme:
\begin{subequations}
\label{Reakt}
\begin{align}
  & A \overset{\rho}{\underset{\kappa}{\rightleftarrows}} 2A \, , \qquad
  A \overset{\sigma}{\rightarrow}\varnothing \, , \label{Reakt1} \\
  & A \overset{\mu}{\rightarrow} B \, , \qquad\
  A + B \overset{\nu}{\rightarrow} B \, ,
\label{Reakt2}
\end{align}
\end{subequations}
supplemented with hopping of the agents $A$ to nearest-neighbors on a 
$d$-dimensional lattice with sites $(i,j,\ldots)$ with diffusion constant 
$\lambda$. The back-reaction in (\ref{Reakt1}) may also be viewed as 
effectively capturing mutual exclusion on the same site. The reactions in
(\ref{Reakt2}) are specific to dIP where the presence of the produced debris 
$B$ suppresses the agent $A$.

The reaction rules (\ref{Reakt}) are reformulated in terms of a master equation
that describes the time dependence of the probability $P(\{n,m\},t)$ for a
given configuration of site occupation numbers $\{n\}=(\ldots,n_i,\ldots)$ and
$\{m\}=(\ldots,m_i,\ldots)$ of the agent $A$ and the debris $B$, respectively.
The configuration probability is encoded in the state vector 
$\left\vert P(t) \right\rangle = \sum_{\{n,m\}} \! P(\{n,m\},t) \, 
\left\vert \{n,m\} \right\rangle$ in a bosonic Fock space spanned by the 
basis $\left\vert \{n,m\} \right\rangle$. These vectors as well as the 
stochastic processes in the master equation are then expressed through the
action of bosonic creation and annihilation operators $\{\hat{a},\hat{b}\}$ and
$\{a,b\}$, respectively, which are defined via 
$\hat{a}_i \left\vert \ldots,n_i,\ldots \right\rangle = 
\left\vert \ldots,n_i+1,\ldots \right\rangle$, and
$a_i \left\vert \ldots,n_i,\ldots \right\rangle = 
n_i \left\vert \ldots,n_i-1,\ldots \right\rangle$, etc. Subsequently, the 
master equation can be written in the form
\begin{equation}
\label{Ham}
  \partial_t \left\vert P(t) \right\rangle = - H \left\vert P(t) \right\rangle
\end{equation}
with an appropriate non-Hermitean pseudo-Hamilton operator 
\cite{Doi76,GrSch80,HoTaVo04}. For example, the reaction scheme (\ref{Reakt}) 
leads to $H = H_{\rm diff} + H_{\rm reac}$ with
\begin{subequations}
\label{H}
\begin{align}
  H_{\rm diff} &= \lambda \sum_{<i,j>} \bigl( \hat{a}_i - \hat{a}_j \bigr) a_i
  = \frac{\lambda}{2} \sum_{<i,j>} \bigl (\hat{a}_i - \hat{a}_j \bigr)
  \bigl(a_i - a_j \bigr) \, , \label{H-diff} \\
  H_{\rm reac} &= \sum_i \Big[ \rho \bigl( 1 - \hat{a}_i \bigr) \hat{a}_i \,
  a_i + \kappa \bigl( \hat{a}_i - 1 \bigr) \hat{a}_i a_i^2 + \sigma \bigl(
  \hat{a}_i - 1 \bigr) a_i \nonumber \\
  &\qquad\qquad + \mu \bigl( \hat{a}_i - \hat{b}_i \bigr) a_i + \nu \bigl( 
  \hat{a}_i - 1 \bigr) \hat{b}_i \, b_i \, a_i \Big] \, . \label{H-react}
\end{align}
\end{subequations}
Here $<i,j>$ denotes a pair of neighboring sites.

In order to compute statistical averages it is necessary to introduce the 
projection state $\left\langle \cdot \right\vert = \left\langle 0 \right\vert 
\prod_i \exp (a_i+b_i)$. Using the identity $\left\langle \cdot \right\vert 
\hat{a}_i = \left\langle \cdot \right\vert = \left\langle \cdot \right\vert 
\hat{b}_i$ one easily finds the expectation value of an observable $A(\{n,m\})$
at time $t$:
\begin{equation}
\label{expv}
  \langle A \rangle(t) = \sum_{\{n,m\}} A(\{n,m\}) \, P(\{n,m\},t) =
  \langle \cdot | A( \{\hat{a} a , \hat{b} b \}) | P(t) \rangle \, ,
\end{equation}
where in the last expression $n$ and $m$ are replaced with the operators 
$\hat{a} a$ and $\hat{b} b$, respectively. The formal solution of the equation 
of motion (\ref{Ham}) reads $\left\vert P(t) \right\rangle = \exp(-t H) 
\left\vert P(0) \right\rangle$. Following standard procedures 
\cite{Pel84,HoTaVo04} the expectation value (\ref{expv}) can be expressed as a 
path integral
\begin{equation}
\label{path}
  \langle A \rangle(t) = \int\mathcal{D}[\hat{a},\hat{b},a,b] \, 
  A(\{ \hat{a} a , \hat{b} b \}) \, \exp (-S[\hat{a},\hat{b},a,b]) \, , 
\end{equation}
with an exponential weight that defines a field theory action. After applying a
(naive) continuum limit, the action becomes
\begin{align}
  S &= \int \! d^dr \, dt \, \Big[ (\hat{a} - 1) \, \partial_t a + \lambda \, 
  \nabla\hat{a} \cdot \nabla a + (\hat{a} - 1) \, (\sigma - \rho \, \hat{a} + 
  \kappa \, \hat{a} a) \, a \nonumber \\
  &\qquad\qquad\qquad + (\hat{b} - 1) \, \partial_t b + \mu (\hat{a} - \hat{b})
  \, a + \nu (\hat{a} - 1) \, \hat{b} b a \Big] \, . \label{S1}
\end{align}
Here, the fields $\hat{a}(\mathbf{r},t)$, $a(\mathbf{r},t)$, 
$\hat{b}(\mathbf{r},t)$, and $b(\mathbf{r},t)$ correspond to the coherent-state
eigenvalues of the bosonic creation and annihilation operators. The 
integrations are subject to $\hat{a} a \geq 0$, $\hat{b} b \geq 0$, and the 
final conditions are $\hat{a}(\mathbf{r}, t \to \infty) = 1 = 
\hat{b}(\mathbf{r},t \to \infty)$. Thus often a field shift according to
$\hat{a} = 1 +\tilde{a}$, $\hat{b} = 1 + \tilde{b}$ is useful. The new 
variables $\tilde{b}$ then appear only linearly in the action (\ref{S1}). 
Hence, they can be integrated out, which results in the differential equation
constraint
\begin{equation}
\label{db}
  \partial_t \, b = (\mu - \nu \, \tilde{a} b) \, a \, ,
\end{equation}
and the new action
\begin{equation}
\label{S2}
  S = \int \! d^dr \, dt \, \tilde{a} \big[ \partial_t - \lambda \nabla^2 
  + (\sigma + \mu - \rho) + (\kappa \, a + \nu \, b - \rho \, \tilde{a}) 
  + \kappa \, \tilde{a} a \big] a \, ,
\end{equation}
where $b$ is given by the solution of Eq.~(\ref{db}). If we now discard all
fourth-order terms which should become irrelevant after the application of
coarse-graining near the critical point, the action (\ref{S2}) attains the
same form as the response functional (\ref{J2}), with the correct sign of the
coupling constants.

Physically, however, the fields are of different origin: Whereas $n$ in the
action $\mathcal{J}$ represents the fluctuating density of the active medium, 
the field $a$ in (\ref{S1}) or (\ref{S2}) results from the bosonic annihilation
operators for the random walks; therefore, $a$ generally is a complex-valued
quantity, and only $\hat{a}$ has to be real and non-negative. Thus, the action 
(\ref{S2}) derived above really is akin to the original Reggeon field theory 
\cite{Gri67,GriMi68,Mo78}. This can be remedied by means of a quasi-canonical 
transformation to proper density variables 
$\hat{a} = 1 + \tilde{a} = \exp(\tilde{n})$, 
$a = n \exp(-\tilde{n})$, $\hat{b} = 1 + \tilde{b} = \exp(\tilde{m})$, and
$b = m \exp(-\tilde{m})$. After integrating by parts the action (\ref{S1}) then
becomes
\begin{align}
  S &= \int \! d^dr \, dt \, \Big\{ \tilde{n} \, \partial_t n + \lambda \big[
  \nabla \tilde{n} \cdot \nabla n - n (\nabla \tilde{n})^2 \big] \nonumber \\
  &\qquad\qquad\quad + \big[ 1 - \exp(-\tilde{n}) \big] 
  \big[ \sigma - \rho \exp(\tilde{n}) + \kappa n\big] n \nonumber \\
  &\qquad\qquad + \tilde{m} \, \partial_t m + \mu \big[ 1 - 
  \exp(\tilde{m}-\tilde{n}) \big] n + \nu \big[ 1 - \exp(-\tilde{n}) \big] 
  m \, n \Big\} \, . \label{S3}
\end{align}
Integration over $\tilde{m}$, followed by the expansion of the exponentials, 
and dispensing with fourth-order terms finally leads to
\begin{align}
  S^{\prime} & = \int \! d^dr \, dt \, \Big\{ \tilde{n} \, \partial_t n
  - \lambda \big[ \tilde{n} \, \nabla^2 n + n (\nabla \tilde{n})^2 \big] 
  \nonumber \\
  &\qquad\qquad\quad + \tilde{n} \big[ (\sigma + \mu - \rho) + \kappa \, n
  + \nu \, m - \frac{\rho+\sigma}{2} \, \tilde{n} \big] n \Big\} \, , 
  \label{S4}
\end{align}
with $\partial_t \, m = \mu \, n$. This action again acquires the same form as 
the response functional (\ref{J2}). Aside from the different meaning of the 
fields, it is reminiscent of the action (\ref{S2}) without the fourth-order 
contributions. However, irrelevant diffusional noise arises in the action 
(\ref{S4}), and the noise term $\propto \tilde{n}^2$ comes with a slightly 
different coupling constant that is invariably negative, even in the ``free'' 
case with $\rho = 0$. Hence, as remarked above, the stochastic processes 
described by the actions $S$ and $S^{\prime}$ are in fact distinct, yet their 
\emph{universal} features are identical. They will however usually differ with 
respect to non-asymptotic, non-universal details.

\subsection{Mean-Field Theory and Naive Scaling Dimensions}

The first approximation in the evaluation of path integrals generally consists 
of a Gaussian truncation in the action $\mathcal{J}$ with respect to the 
fluctuations about the maximum of the statistical weight $\exp(-\mathcal{J})$. 
The extrema are in turn determined from the saddle-point (mean-field) equations
\begin{subequations}
\label{SPE}
\begin{align}
  0 = \frac{\delta\mathcal{J}}{\delta\tilde{n}} &= \Big[ \partial_t + \lambda
  (\tau - \nabla^2) + \lambda (g_1 \, n + g_2 \, m - 2 g_3 \, \tilde{n}) \Big]
  n - q \, , \label{SPE1} \\
  0 = \frac{\delta\mathcal{J}}{\delta n} &= \Big[ - \partial_t + \lambda
  (\tau - \nabla^2) + \lambda (2 g_1 \, n + g_2 \, m - g_3 \, \tilde{n}) \Big]
  \tilde{n} \nonumber \\
  &\qquad\qquad\qquad\qquad + \lambda \, g_2 \int_t^\infty \! dt^{\prime} \, 
  n(t^{\prime}) \, \tilde{n}(t^{\prime}) \, . \label{SPE2}
\end{align}
\end{subequations}
For $t \to \infty$, the stable homogeneous stationary solutions in the case of 
homogeneous vanishing source $q = 0$ are $\tilde{n} = 0$, and $n = 0$ if 
$\tau > 0$. For $\tau < 0$ we obtain $n = |\tau|/g_1$ for $g_2 = 0$ (DP), and 
$n = 0$, $m = |\tau|/g_2$ for $g_1 = 0$ (dIP). This demonstrates that within 
the mean-field approximation the critical point is located at $\tau = 0$, and 
in its vicinity on the active side the order parameter vanishes as 
$n \sim |\tau|^\beta$ with critical exponent $\beta = 1$.

Now let us scale spatial distances $x$ by a convenient mesoscopic length scale 
$\mu^{-1} \gg \Lambda^{-1}$. Consequently we find the naive scaling dimensions 
$\lambda t \sim \mu^{-2}$, $\tau \sim \mu^2$, and $g_2/g_1 \sim \mu^2$. Hence, 
in the asymptotic long-time and large-distance limit, $\tau$ and $g_2/g_1$ 
constitute \emph{relevant} parameters, flowing to $\infty$ under successive RG
scale transformations. The last relation shows that the coupling $g_1$ becomes 
irrelevant, provided $g_2 > 0$ (dIP). Since the action $\mathcal{J}$ is 
dimensionless, we find from Eq.~(\ref{J2}) that $\tilde {n} n\sim \mu^d$. It is
characteristic of the field theory for spreading phenomena with an absorbing 
state that the dynamic response functional $\mathcal{J}$ contains a 
\emph{redundant} parameter \cite{We74} that must be eliminated by a suitable 
rescaling
\begin{equation}
\label{RedScal}
  \tilde{n} = K^{-1} \tilde{s} \, , \quad n = K s \, , \quad m = K S \, , 
\end{equation}
with an amplitude $K$ that generally carries nonvanishing scaling dimension. 
Because both the lowest-order non-vanishing coupling constants in the 
mean-field part of the action and the stochastic noise strength are clearly
required for a meaningful non-trivial perturbation expansion, it is convenient 
to choose $K$ such that the corresponding couplings attain the same scaling 
dimensions. Thus we set $2 K g_1 = 2 K^{-1} g_3 = g$ in the case of DP, and 
$K g_2 = 2 K^{-1} g_3 = g$ for dIP, where $K^{-1}$ parametrizes, e.g., the 
line of DP transitions in the Domany-Kinzel automaton \cite{DoKi84}. We note
that this amplitude characterizes the non-universal ``lacunarity'' property of 
percolating clusters which tends to zero in the case of compact percolation. 
The dynamic response functional now assumes the two distinct forms
\begin{subequations}
\label{J3}
\begin{align}
  \mathcal{J}_{\rm DP} & = \int \! d^dr \, dt \, \Big\{ \tilde{s} \big[
  \partial_t + \lambda (\tau - \nabla^2) + \frac{\lambda g}{2} \, 
  (s - \tilde{s}) \big] s - \lambda h \, \tilde{s} \Big\} \, , \label{JDP} \\
  \mathcal{J}_{\rm dIP} & = \int \! d^dr \, dt \, \Big\{ \tilde{s} \Big[
  \partial_t + \lambda (\tau - \nabla^2) + \frac{\lambda g}{2} \,
  (2S - \tilde{s}) \big] s - \lambda h \, \tilde{s} \Big\} \, , \label{JdIP}
\end{align}
\end{subequations}
for DP and dIP, respectively, where $\lambda h = K^{-1} q$. Hence, a seed at 
the origin, i.e., a sick individual at $(\mathbf{r},t) = (\mathbf{0},0)$, is 
represented by $\lambda h(\mathbf{r},t) = K^{-1} \delta(\mathbf{r}) \delta(t)$.
After fixing the redundancy in this manner the naive scaling of the fields and 
couplings is uniquely given by
\begin{subequations}
\label{Skal}
\begin{align}
  \text{DP:} \qquad & s \sim \tilde{s} \sim\mu^{d/2} \, , \quad 
  g \sim \mu^{(4-d)/2} \, , \label{DP-Skal} \\
  \text{dIP:} \qquad & S \sim \tilde{s} \sim \mu^{(d-2)/2} \, , \quad 
  s \sim \mu^{(d+2)/2} \, , \quad g \sim \mu^{(6-d)/2} \, . \label{dIP-Skal}
\end{align}
\end{subequations}

From the scaling dimensions of the couplings $g$ we infer the upper critical 
dimensions $d_c = 4$ for DP, and $d_c = 6$ for dIP. At this point it is also
straightforward to show that all higher-order terms in the gradient and density
expansions that we had neglected before in fact acquire negative scaling 
dimensions near $d_c$. This proves that those terms are indeed irrelevant for
the asymptotic IR scaling behavior. Note that relevance and irrelevance are 
assigned here with reference to the Gaussian theory. After discarding the 
irrelevant terms, the dynamic response functionals (\ref{Skal}) display duality
invariance with respect to time inversion,
\begin{equation}
\label{ZeitSp}
  \text{DP:} \quad \tilde{s}(t) \leftrightarrow - s(-t) \, , \qquad
  \text{dIP:} \quad \tilde{s}(t) \leftrightarrow - S(-t) \, .
\end{equation}
However, in general these symmetries only hold asymptotically.

\subsection{IR Problems and Renormalization}

The important goal of statistical theories is the determination of correlation
and response functions (generally called Green's functions) of the dynamical
variables as functions of their space-time coordinates, as well as of the 
relevant control parameters. In a compact form, one attempts to determine the 
\emph{cumulant generating functional}
\begin{equation}
\label{KumGen}
  \mathcal{W}[H,\tilde{H}] = \ln \int \mathcal{D}[\tilde{s},s] \, \exp
  \Big[ - \mathcal{J}[\tilde{s},s] + (H,s) + (\tilde{H},\tilde{s}) \Big] \, .
\end{equation}
Functional derivatives with respect to the sources $H$ and $\tilde{H}$,
\begin{equation}
\label{DerKum}
  \left. \frac{\delta^{N+\tilde{N}} \, \mathcal{W}}{[\delta H^N]
  [\delta\tilde{H}^{\tilde{N}}]} \right\vert_{H=\tilde{H}=0} = \langle 
  [s^N][\tilde{s}^{\tilde{N}}] \rangle^{\rm (cum)} =: G_{N,\tilde{N}} \, ,
\end{equation}
define the \emph{Green's functions} (here we suppress all space and time 
coordinates). In general, an exact expression for $\mathcal{W}[H,\tilde{H}]$
cannot be found and one must resort to a perturbational evaluation. 
Perturbation theory is developed starting from the Gaussian contribution 
$\exp(-\mathcal{J}_0)$ to the weight, and the subsequent expansion of the
remainder $\exp(-\mathcal{J}_i)$, where 
$\mathcal{J}_i = \mathcal{J}-\mathcal{J}_0$. 

The different contributions to the perturbation series are graphically 
organized in successive order of closed loops in terms of linked diagrams 
which, in translationally invariant theories, can be decomposed into one-line 
irreducible amputated Feynman diagrams that represent the building blocks for 
the \emph{vertex functions}. The generating functional for the vertex functions
$\Gamma[\tilde{s},s]$ is related to the cumulant generating functional via the 
Legendre transformation
\begin{subequations}
\label{Vert-Kum}
\begin{align}
  \Gamma\lbrack\tilde{s},s] & +\mathcal{W}[H,\tilde{H}] 
  = (H,s) + (\tilde {H},\tilde{s}) \, , \label{Vert-Kum-1} \\
  \text{with} \ &\ s =\frac{\delta\mathcal{W}}{\delta H} \, , \quad
  \tilde{s} = \frac{\delta\mathcal{W}}{\delta\tilde{H}} \, , \label{Vert_Kum-2}
\end{align}
\end{subequations}
and vice versa. It is easy to see that to zero-loop (`tree') order 
$\Gamma[\tilde{s},s]$ is identical to $\mathcal{J}[\tilde{s},s]$. The vertex 
functions are defined via the functional derivatives
\begin{equation}
\label{DerVert}
  \left. \frac{\delta^{N+\tilde{N}} \Gamma[\tilde{s},s]}
  {[\delta^{\tilde{N}}\tilde{s}][\delta s^N]} \right\vert_{\tilde{s}=s=0} =: 
  \Gamma_{\tilde{N},N} \, , 
\end{equation}
and they are represented by irreducible diagrams if $\Gamma_{1,0} = 0$. Note
that all the $\Gamma_{0,N}$ necessarily vanish as a consequence of causality. 
The Gaussian parts of the response functionals Eqs.~(\ref{J3}) define the 
\emph{propagator}
\begin{subequations}
\label{Prop}
\begin{align}
  \langle s(\mathbf{r},t) &\tilde{s}(\mathbf{r}^{\prime},t^{\prime}) \rangle_0
  =: G(\mathbf{r}-\mathbf{r}^{\prime},t-t^{\prime}) \, , \label{Prop1}\\
  G(\mathbf{r},t) & = \int_{\mathbf{q},\omega} \frac{\exp(i\mathbf{q \cdot r} 
  - i\omega t)}{- i \omega + \lambda(\mathbf{q}^2 + \tau)} \, , \label{Prop2}
\end{align}
where the momentum integral is limited by the ultraviolet (UV) cutoff 
$\Lambda$. We have within the tree approximation
\end{subequations}
\begin{equation}
\label{Zero-Loop}
  \Gamma_{1,1}(\mathbf{q},\omega) = G_{1,1}(\mathbf{q},-\omega)^{-1} =
  i \omega + \lambda (\mathbf{q}^{2} + \tau) + \text{loop corrections} \, .
\end{equation}
The non-linear contributions from $\mathcal{J}_i$ yield the vertices of the 
field theory.

As is common with quantum and statistical field theories, the naive 
perturbation expansion, here based on our response functionals (\ref{J3}), is 
problematic for several reasons 
\cite{Am84,ZiJu96,Sy73,SchlDo89,BaBe90,Do94,Schae94}. To be specific, consider 
the inverse susceptibility $\chi(\tau,g,\Lambda,\varepsilon)^{-1} = 
\lambda^{-1} \Gamma_{1,1}(\mathbf{q}=0,\omega=0) = \tau + O(g^2)$ that 
describes the linear response to a stationary homogeneous external source. We
shall be interested in how the structure of the theory depends on the spatial 
dimension $d$ as encoded in the parameter $\varepsilon = d_c-d$. First, the 
perturbation series in $g^2$ in general is divergent. Although it is 
characterized by a vanishing radius of convergence, it presumably constitutes 
an asymptotic series and should be resummable. Yet more severe problems arise 
from the massless critical limit. The critical point $\tau_c$, defined 
implicitly by $\chi(\tau_c,g,\Lambda,\varepsilon)^{-1} = 0$, may by dimensional
arguments be written in the form
\begin{equation}
\label{tau_c}
  \tau_c = g^{4/\varepsilon} \, S(g\Lambda^{-\varepsilon/2},\varepsilon) \, .
\end{equation}
However, the function $S(z,\varepsilon)$ is not calculable by means of 
perturbation theory, since it diverges in the limit $z \to 0$ for all positive
$\varepsilon \leq 2$. The limit $S(0,\varepsilon) =: S(\varepsilon)$ does exist
for $\varepsilon > 2$, and an analytic continuation to $\varepsilon \leq 2$ 
defines the dimensionally regularized Symanzik function $S(\varepsilon)$. But 
now $S(\varepsilon)$ displays simple IR poles at all $\varepsilon = 2/n$ with
positive integers $n$. Note that at these discrete points 
$\tau_c \propto g^{2n}$.

Let us now introduce the new variable $\hat{\tau} = \tau - \tau_c$. The vertex 
functions $\Gamma_{\tilde{N},N}(\{\mathbf{q},\omega\},\hat{\tau})$, regarded as
functions of $\hat{\tau}$ instead of $\tau$, are free of IR singularities for 
$\hat{\tau} \not= 0$, but they require the perturbationally inaccessible 
function (\ref{tau_c}). In order to eliminate the IR poles completely from the 
calculation in any dimension $d < d_c$, it is possible at this stage to change 
variables to the correlation length squared 
\cite{SchlDo89},
\begin{equation}
\label{Korr-L}
  \xi(\hat{\tau},g,\Lambda)^2 = \frac{\partial}{\partial\mathbf{q}^2} \, 
  \ln \Gamma_{1,1}(\mathbf{q},\omega) \bigg\vert _{\mathbf{q}^2=\omega=0} \, .
\end{equation}
Since the function $\xi$, which does not display IR poles either, should be a 
monotonic function of $\hat{\tau}$, it can be inverted to 
$\hat{\tau}=\hat{\tau}(\xi,g,\Lambda)$ with $\hat{\tau}(\infty,g,\Lambda) = 0$.
Finally one may substitute $\hat{\tau}$ by $\hat{\tau}(\xi,g,\Lambda)$ and thus
arrive at vertex functions written in terms of $\xi$, i.e., 
$\hat{\Gamma}_{\tilde{N},N}(\{\mathbf{q},\omega\},\xi,g,\Lambda) =
\Gamma_{\tilde{N},N}(\{\mathbf{q},\omega\},\hat{\tau}(\xi,g,\Lambda),g,
\Lambda)$, which are now calculable in perturbation theory and are devoid of IR
singularities for $\varepsilon > 0$ and $\xi < \infty$.

We return to the consideration of the analytical properties of the response 
function. Dimensional analysis yields
\begin{equation}
\label{F}
  \chi(\tau,g,\Lambda)^{-1} = \hat{\tau} \,
  F(\hat{\tau}\Lambda^{-2},g\Lambda^{-\varepsilon/2},\varepsilon) \, ,
\end{equation}
with the perturbational expansion
\begin{equation}
\label{F-pert}
  F(\theta,z,\varepsilon) = 1 + \sum_{n=1}^\infty f_n(\theta,\varepsilon) \,
  z^{2n} \, .
\end{equation}
For $\varepsilon > 0$, the functions $f_n$ are divergent in the critical limit 
$\theta \to 0$, a direct consequence of the fact that 
$\chi^{-1} \sim |\hat{\tau}|^\gamma$ for $\hat{\tau} \to 0$ with a critical 
exponent $\gamma \neq 1$ if $d < d_c$. Hence, a series in $\ln \theta$ appears 
to be produced by the perturbational expansion which needs to be properly 
resummed.

Wilson's momentum-shell renormalization procedure \cite{Wils75} fully avoids 
the IR problems of the naive perturbation expansion. Furthermore, in contrast 
to renormalized field theory, Wilson's approach does not require the 
elimination of the IR-irrelevant couplings prior to deriving the RG flow
equations, and so pre-asymptotic critical as well as crossover properties are 
calculable. However, even though Wilson's RG procedure in that sense is 
superior to the field-theoretic method, it is generally not advisable for a 
systematic calculation of universal properties to higher then one-loop order. 
In contrast, the field-theoretic RG method \cite{Am84,ZiJu96} does not proceed 
by successive elimination of short-wavelength degrees of freedom and a 
rescaling of parameters, but instead exploits the UV-renormalizability of the 
perturbation expansion for $d \leq d_c$ to enable a mapping from the critical 
to a non-critical region in parameter space wherein perturbational calculations
are unproblematic. Both methods are fully equivalent with respect to describing
asymptotic and universal features.

We rename the original bare fields and parameters according to 
$s \to \mathring{s}$, $\tilde{s} \to \mathring{\tilde{s}}$, 
$\tau \to \mathring{\tau}$, etc. In accord with the symmetries (\ref{ZeitSp})
we choose the following multiplicative renormalizations
\begin{subequations}
\label{Ren-Z}
\begin{align}
  \mathring{s} &= Z^{1/2} s \, , \qquad\qquad\quad\,\
  \mathring{\tilde{s}} = \tilde{Z}^{1/2} \tilde{s} \, , \qquad\quad\,\
  G_{\varepsilon} \mathring{g}^2 = \tilde{Z}^{-1} Z_\lambda^{-2} Z_u u
  \mu^{\varepsilon} \, , \label{Ren-Z1}\\
  \mathring{\lambda} &= (Z \tilde{Z})^{-1/2} Z_\lambda \lambda \, , \qquad
  \mathring{\tau} = Z_\lambda^{-1} Z_\tau \tau + \mathring{\tau}_c \, , \qquad
  \mathring{h} = Z^{1/2} Z_\lambda^{-1} h \, , \label{Ren-Z2}\\
  \tilde{Z} &= Z \quad \text{for DP} \, , \qquad\,\ \
  \tilde{Z} = Z_\lambda \quad\text{for dIP} \, , \label{Ren-Z3}
\end{align}
\end{subequations}
where $G_{\varepsilon} = \Gamma(1+\varepsilon/2)/(4\pi)^{d/2}$ is a convenient
amplitude. $u$ represents the dimensionless coupling constant, and $\tau = 0$ 
at the critical point. The renormalization constants 
$Z_{\cdots} = Z_{\cdots}(u,\mu/\Lambda,\varepsilon)$ can be chosen in a 
UV-renormalizable theory in such a way that
\begin{subequations}
\label{Ren-GG}
\begin{align}
  &\mathring{\Gamma}_{\tilde{N},N}(\{\mathbf{q},\omega\},\mathring{\tau},
  \mathring{g},\mathring{\lambda},\Lambda) = 
  \frac{1}{\tilde{Z}^{\tilde{N}/2} Z^{N/2}}
  \Gamma_{\tilde{N},N}(\{\mathbf{q},\omega\},\tau,u,\lambda,\mu) 
  \Big[ 1 + O\big( (\Lambda\xi)^{-\Delta} \big) \Big] , \label{Ren-Gamma}\\
  &\mathring{G}_{N,\tilde{N}}(\{\mathbf{r},t\},\mathring{\tau},
  \mathring{g},\mathring{\lambda},\Lambda) = Z^{N/2} \tilde{Z}^{\tilde{N}/2}
  G_{N,\tilde{N}}(\{\mathbf{r},t\},\tau,u,\lambda,\mu) 
  \Big[ 1 + O\big( (\Lambda\xi)^{-\Delta} \big) \Big] , \label{Ren-Greens}
\end{align}
\end{subequations}
with $\Delta = 2 + O(\varepsilon)$. Within each successive order of the 
perturbation expansion, the renormalized vertex functions 
$\Gamma_{\tilde{N},N}$ are thereby rendered finite and well-defined. Note
that the physical bare and the renormalized vertex function display the same
dependence on the variables $(\mathbf{q},\omega,\tau)$ in the limit 
$(\Lambda\xi) \to \infty$ up to nonuniversal amplitudes. In principle, the
infinite cutoff limit is unphysical and only employed here to develop a 
systematic RG mapping that works effectively to higher orders in the loop 
expansion. The theory becomes only UV renormalizable at the upper critical 
dimension $d_c$ (super-renormalizable below $d_c$) because we have previously
eliminated all IR-irrelevant couplings and shifted the relevant control 
parameters (here, $\tau$) to be zero at the critical point. Indeed, the 
problematic UV and IR singularities are linked precisely at the critical 
dimension $d = d_c$. For $d > d_c$ the field theory is IR-finite and 
UV-infinite, whereas conversely for $d < d_c$ it is IR-infinite and UV-finite.

For $d \leq d_c$, the renormalization factors diverge in three distinct limits
The first two represent UV divergences, while the third one constitutes an 
IR singularity:
$\ln Z_{\cdots}(u,\mu/\Lambda,\varepsilon) \to \infty$ \quad if
\begin{enumerate}
\item $\mu/\Lambda \to 0$, $\varepsilon = 0$, $u$ fixed;
\item $\mu/\Lambda = 0$, $\varepsilon \to 0$, $u$ fixed;
\item $\mu/\Lambda \to l \mu / \Lambda \overset{l \to 0}{\longrightarrow} 0$, 
      $u \to u(l) \overset{l \to 0}{\longrightarrow}u_{\ast}$, 
      $\varepsilon > 0$ fixed.
\end{enumerate}
Here $u_{\ast}$ denotes the first nontrivial zero of the Gell-Mann--Low 
function $\beta(u)$ to be introduced below, and $u(l)$ is the solution of the 
RG flow equation $l \, du(l) / dl = \beta(u(l))$. Hence, the determination of 
the $Z$ factors from the UV divergences provides us at the same time with 
important information on the critical IR singularities and thereby on critical 
exponents. This observation lies at the heart of the field-theoretic RG method.
Explicit calculations of the renormalization constants are facilitated if one 
first takes the continuum limit $\Lambda \to \infty$ with $\varepsilon > 0$ 
(\emph{dimensional regularization}) together with the requirement that the $Z$
factors absorb just the $\varepsilon$ poles (\emph{minimal subtraction}). 
Notice that this minimal dimensional regularization does not at all require
an $\varepsilon$ expansion.

Yet the continuum limit raises subtle problems in the realm of statistical 
physics \cite{BaBe90,Schae94}. For example, one may conclude from the above
remarks that the full range from zero to infinity of the bare coupling constant
$\mathring{g}$ is mapped only to the interval $[0,u_{\ast}]$ which lies
between the UV-stable fixed point $u = 0$ and the IR-stable fixed point
$u = u_{\ast}$. Values of $u$ larger than $u_{\ast}$ are excluded. On the other
hand, there exist many pre-asymptotic simulational and experimental results,
e.g., for the three-dimensional Ising model near the critical point and some
polymer systems, which are linked to this strong-coupling region that does not
include an UV-stable fixed point. In Wilson's RG approach this regime can be
reached by a choosing suitable initial values of the relevant and irrelevant 
couplings, and a finite UV cutoff $\Lambda$. The UV-renormalized field theory 
should therefore with $\Lambda \to \infty$ should therefore be accepted as an 
\emph{effective} field theory where the influences of irrelevant couplings and 
the finite cutoff are implicitly encoded in the full range of the coupling $u$ 
also above the fixed point.

\subsection{RG Flow Equations and Critical Exponents}

Next, we compare renormalized vertex functions from the same bare theory, but 
which are renormalized at two different external momentum scales $\mu$ and 
$l \mu$, respectively. The bare vertex functions are held fixed, and we shall 
henceforth neglect the $O\big( (\Lambda\xi)^{-\Delta} \big)$ corrections. Then 
it follows from Eq.~(\ref{Ren-GG}) that
\begin{subequations}
\label{Mapping}
\begin{align}
  \Gamma_{\tilde{N},N}(\{\mathbf{q},\omega\},\tau,u,\lambda,\mu) &=
  X_{\tilde{N},N}\big( u,u(l) \big)^{-1} \, \Gamma_{\tilde{N},N}\big( 
  \{\mathbf{q},\omega\},\tau(l),u(l),\lambda(l),l\mu \big) \, , \label{Mapp1}\\
  G_{N,\tilde{N}}(\{\mathbf{r},t\},\tau,u,\lambda,\mu) &= 
  X_{\tilde{N},N}\big( u,u(l) \big) \, G_{N,\tilde{N}}\big( 
  \{\mathbf{r},t\},\tau(l),u(l),\lambda(l),l\mu \big) \, ,
\label{Mapp2}
\end{align}
\end{subequations}
where $X_{\tilde{N},N} = \tilde{X}^{\tilde{N}/2} X^{N/2}$ and
\begin{equation}
\label{Mapp-Fu}
  \tilde{X}\big( u(l),u \big) = \lim_{\Lambda \to \infty} \frac{\tilde{Z}\big( 
  u(l),l\mu/\Lambda,\varepsilon \big)}{\tilde{Z}(u,\mu/\Lambda,\varepsilon)} \,
  , \quad X\big( u(l),u \big) = \lim_{\Lambda \to \infty} \frac{Z\big(
  u(l),l\mu/\Lambda,\varepsilon \big)}{Z(u,\mu/\Lambda,\varepsilon)} \, .
\end{equation}
This is the desired mapping. If $\tau/\mu\ll1$, the left-hand site of
Eq.~(\ref{Mapping}) is IR-problematic, whereas for an appropriate $l\ll1$ the
vertex function on the right-hand site is unproblematic if we choose
$\tau(l)/(l\mu)^{2}\approx1$. It is expected that $\lim_{l\rightarrow
0}u(l)=u_{\ast}$, and the leading critical properties are transformed to the
functions $\tilde{X}$ and $X$.

Casting this mapping in differential form yields the \emph{renormalization 
group equation} (RGE). We let $\Lambda \to \infty$ and define the RG functions 
as logarithmic derivatives with respect to the normalization scale $\mu$,
holding bare parameters fixed,
\begin{subequations}
\label{RG-Fu}
\begin{align}
  \beta(u) &= \frac{\partial u}{\partial \ln\mu}\bigg\vert_0 \, , \qquad
  \gamma(u) = \frac{\partial \ln Z}{\partial \ln\mu}\bigg\vert_0 \, , \qquad
  \tilde{\gamma}(u) = \frac{\partial \ln\tilde{Z}}{\partial \ln\mu}\bigg\vert_0
  \, , \label{RG-Fu1}\\
  \kappa(u) &= \frac{\partial \ln\tau}{\partial \ln\mu} \bigg\vert_0 \, , \,
  \qquad \zeta(u) = \frac{\partial \ln\lambda}{\partial \ln\mu}\bigg\vert_0 \, 
  . \label{RG-Fu2}
\end{align}
\end{subequations}
Consider the renormalized Green's functions
\begin{equation}
\label{Ren-Green}
  G_{N,\tilde{N}}(\{\mathbf{r},t\},\tau,u,\lambda,\mu) = 
  \lim_{\Lambda \to \infty} \big[ Z^{-N/2} \tilde{Z}^{-\tilde{N}/2}
  \mathring{G}_{N,\tilde{N}}(\{\mathbf{r},t\},\mathring{\tau},\mathring{g},
  \mathring{\lambda},\Lambda) \big] \, ;
\end{equation}
using the definitions (\ref{RG-Fu}) it is straightforward to derive the RGE
\begin{equation}
\label{RGE}
   \Big[ \mu \frac{\partial}{\partial\mu} + \zeta \lambda 
   \frac{\partial}{\partial\lambda} + \kappa \tau \frac{\partial}{\partial\tau}
   + \beta\frac{\partial}{\partial u} + \frac{1}{2} (N \gamma + \tilde{N}
   \tilde{\gamma}) \Big] \, 
   G_{N,\tilde{N}}(\{\mathbf{r},t\},\tau,u,\lambda,\mu) = 0 \, .
\end{equation}
The solution of this partial differential equation is provided by the method of
characteristics, which solve the ordinary differential equations
\begin{subequations}
\label{Char}
\begin{align}
  \bar{\mu}(l) &= l \mu \, , \qquad\ \ l \frac{d\ln X}{dl} = \gamma(\bar{u})
  \, , \qquad\!\! l \frac{d\ln\tilde{X}}{dl} = \tilde{\gamma}(\bar{u}) \, ,
  \label{Char1}\\
  l \frac{d\bar{u}}{dl} &= \beta(\bar{u}) \, , \qquad 
  l \frac{d\ln\bar{\tau}}{dl} = \kappa(\bar{u}) \, , \qquad 
  l \frac{d\ln\bar{\lambda}}{dl} = \zeta(\bar{u}) \, , \label{Char2}
\end{align}
\end{subequations}
with the initial conditions $X(1) = 1 = \tilde{X}(1)$, $\bar{u}(1) = u$,
$\bar{\tau}(1) = \tau$, and $\bar{\lambda}(1) = \lambda$.
This yields the general expression
\begin{equation}
\label{RG-Sol}
  G_{N,\tilde{N}}(\{\mathbf{r},t\},\tau,u,\lambda,\mu) = X(l)^{N/2}
  \tilde{X}(l)^{\tilde{N}/2} \, G_{N,\tilde{N}}(\{\mathbf{r},t\},\bar{\tau}(l),
  \bar{u}(l),\bar{\lambda}(l),l\mu) \, .
\end{equation}

In the critical limit $l \to 0$ the characteristic equations provide us with
the \emph{asymptotic scaling laws}. From Eqs.~(\ref{Char}) we learn that 
$\bar{u}(l)$ flows to a stable fixed point $u_{\ast}$, given as a zero of the 
RG beta function $\beta(u_{\ast}) = 0$, provided the first derivative 
$\beta'((u_{\ast}) > 0$. With the definitions
\begin{equation}
\label{FP-Werte}
  \eta = \gamma(u_{\ast}) \, , \qquad \tilde{\eta} = \tilde{\gamma}(u_{\ast})
  \, , \qquad \nu^{-1} = 2 - \kappa(u_{\ast}) \, , \qquad 
  z = 2 + \zeta(u_{\ast}) \, ,
\end{equation}
and employing dimensional scaling, we obtain from Eq.~(\ref{RG-Sol}) the
asymptotic form for the Green's functions 
\begin{equation}
\label{As-Skal}
  G_{N,\tilde{N}}^{(\rm as)}(\{\mathbf{r},t\},\tau,u,\lambda,\mu) = 
  A^N \tilde{A}^{\tilde{N}} \, l^{\delta_{N,\tilde{N}}} \, F_{N,\tilde{N}}\big(
  \{l \mu \mathbf{r},l^z A_\lambda \lambda \mu^2 t\},l^{-1/\nu} A_\tau 
  \tau/\mu^2 \big) \, .
\end{equation}
Here, $A$, $\tilde{A}$, $A_\lambda$, and $A_\tau$ represent four 
\emph{non-universal} amplitudes. The symmetries (\ref{ZeitSp}) provide us with 
\emph{scaling relations} between the critical exponents. For DP, 
Eq.~(\ref{Ren-Z3}) implies $\gamma = \tilde{\gamma}$, whereas 
$\gamma = \tilde{\gamma} + 2 \zeta$ for dIP. Thus we obtain 
\begin{equation}
\label{Skalenrel}
  \tilde{\eta} = \eta \quad \text{for DP} \, , \qquad 
  2 (z-2) = \eta - \tilde{\eta} \quad \text{for dIP} \, .
\end{equation}
The scaling functions $F_{N,\tilde{N}}$ are universal, and the exponents
$\delta_{N,\tilde{N}}$ are given by
\begin{subequations}
\label{Delta}
\begin{align}
  \delta_{N,\tilde{N}} &= \frac{\beta}{\nu}(N+\tilde{N}) \, , \qquad\qquad\ \
  \beta = \nu \, \frac{d + \eta}{2} \qquad\quad\ \ \text{for DP} 
  \label{Delta-DP}\\
  \delta_{N,\tilde{N}} &= \frac{\beta}{\nu}(N+\tilde{N}) + zN \, , \qquad
  \beta = \nu \, \frac{d - 2 +\tilde{\eta}}{2} \qquad \text{for dIP} \, . 
  \label{Delta-dIP}
\end{align}
\end{subequations}
Consequently, there are only \emph{three} independent scaling exponents $\eta$,
$z$, and $\nu$. In principle, a similar reduction may be found also for the
number of independent non-universal amplitudes. However, the rescaling 
(\ref{RedScal}) introduces an additional amplitude $K$ related to $A$ and 
$\tilde{A}$ by $K^2 = A /\tilde{A}$.

In summary, we have established that percolation processes near the critical 
point are asymptotically described by universal scaling functions with three
scaling exponents, but four non-universal amplitudes. The remaining task is to
explicitly calculate the $Z$ factors and the universal scaling functions in the
non-critical region by means of perturbation theory. There are two methods to
avoid the non-calculable Symanzik function. Either one changes variables from 
the parameter $\tau$ to the correlation length $\xi$, see Eq.~(\ref{Korr-L}), 
which is invariant under renormalization: 
$\mathring{\xi}(\mathring{\tau},\mathring{g},\Lambda) = \xi(\tau,u,\mu)$, and 
then calculates perturbationally the derivative of $\tau$ by $\xi^{-2}$ which 
eliminates $\tau_c$ \cite{SchlDo89}. Alternatively, one may apply, in addition 
to the perturbation expansion, a dimensional $\varepsilon$ expansion which 
formally sets $\tau_c$ to zero, see Eq.~(\ref{tau_c}).

\section{Field Theory of Directed Percolation}

Directed percolation constitutes perhaps the simplest model of a strictly
non-equilibrium system that displays a continuous phase transition. In this 
respect its role is comparable with the paradigmatic Ising model for 
equilibrium critical phenomena. Correspondingly, the DP field theory as given 
by the response functional (\ref{JDP}) can be regarded as the non-equilibrium 
analog of the $\phi^4$ field theory. In the following sections we shall 
consider the renormalized DP field theory and the ensuing asymptotic scaling of
important quantities in detail.

\subsection{Perturbation Theory, Renormalization, and Asymptotic Scaling}

\begin{figure}[ptb]
\begin{center}\includegraphics[width=5cm]{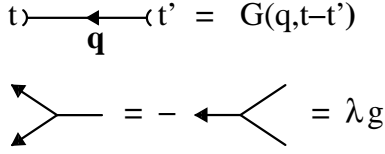}\end{center}
\caption{Elements of the diagrammatic perturbation expansion: propagator (top) 
         and three-point vertices (bottom).}
\label{elemente}
\end{figure}

As stated before, the perturbation expansion is arranged loop-wise with respect
to the harmonic part of the response functional (\ref{JDP}). In the 
momentum-time representation, the propagator (\ref{Prop2}) reads
\begin{equation}
\label{Prop(t)}
  \mathring{G}(\mathbf{q},t) = \theta(t) \, \exp\bigl[ - \mathring{\lambda}
  (\mathring{\tau} + q^2) t \bigr] \, ,
\end{equation}
where the Heaviside step function is defined with $\theta(t=0) = 0$. This
follows from the It\^o discretization of the path integral and ensures 
causality. The anharmonic coupling terms in $\mathcal{J}_{\rm DP}$ define the
elements of the graphical perturbation expansion, depicted in 
FIG.~\ref{elemente}: An arrow marks a $\tilde{s}$-``leg'', and we 
conventionally draw diagrams with the arrowss always directed to the left 
(i.e., we employ ascending time ordering from right to left). The perturbation 
series of the translationally invariant field theory can be analyzed through
calculation of the vertex functions 
$\Gamma_{\tilde{N},N}(\left\{ \mathbf{q},\omega\right\})$ that correspond to 
the one-particle irreducible ``amputated'' graphs.

\begin{figure}[b]
\begin{center}\includegraphics[width=9cm]{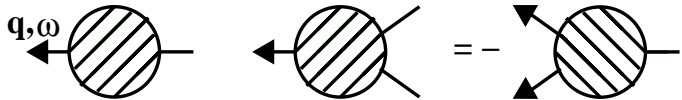}\end{center}
\caption{The naively divergent vertex functions $\Gamma_{1,1}$ and 
     $\Gamma_{1,2} = - \Gamma_{2,1}$ in the DP field theory.}
\label{vertfu}
\end{figure}

We are now ready to consider the renormalization of the DP field theory. To
this end, we evaluate all diagrammatic contributions to the vertex functions to
a given loop order by means of dimensional regularization, and subsequently 
absorb the UV divergences order by order in the loop expansion into a suitable 
renormalization of the fields and model parameters. We choose the scheme 
(\ref{Ren-Z}) and determine the renormalization constants in the minimal 
subtraction prescription. Absorbing the $\varepsilon$ poles of the naively 
divergent vertex functions into Z factors in fact renormalizes the full theory.
The naively divergent vertex functions carry non-negative scaling dimensions. 
In the case of DP, these are $\Gamma_{1,1} \sim \mu^2$ and 
$\Gamma_{1,2} = - \Gamma_{2,1} \sim \mu^\varepsilon$, schematically illustrated
in FIG.~\ref{vertfu}.

\begin{figure}[ptb]
\begin{center}\includegraphics[width=6.5cm]{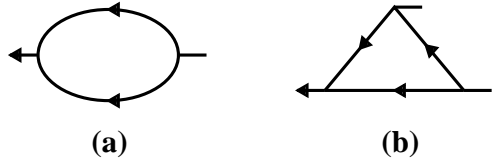}\end{center}
\caption{The one-loop Feynman diagrams for DP.}
\label{1-loop}
\end{figure}

We will now explicitly determine the renormalizations to one-loop order. The
primitively divergent one-loop graphs are shown in FIG.~\ref{1-loop}. We begin
by expressing the one-loop contribution to the propagator self-energy, 
FIG.~\ref{1-loop}(a), as a function of external momentum $\mathbf{q}$
and frequency $\omega$:
\begin{align}
\label{3(a)}
  6(a) &= - \frac{\mathring{\lambda} \mathring{g}^2}{2} \int_\mathbf{p}
  \frac{1}{i\omega / \mathring{\lambda} + 2 \mathring{\tau} +
  (\mathbf{p-q}/2)^2 + (\mathbf{p+q}/2)^2} \nonumber \\
  &= \frac{G_\varepsilon}{2 \varepsilon} \, \mathring{\tau}^{-\varepsilon/2}
  \, \mathring{\lambda} \mathring{g}^{\,2} \biggl( 
  \frac{2 \mathring{\tau}}{2-\varepsilon} + \frac{i\omega}{2\mathring{\lambda}}
  + \frac{\mathbf{q}^2}{4} \biggr) + \ldots \, . 
\end{align}
Here we have retained only terms linear in $\omega$ and $\mathbf{q}^2$. These 
are the contributions that display poles in $\varepsilon = 4-d$. The first term
in the brackets has an IR singularity at $\varepsilon = 2$ ($d = 2$). This pole
can be eliminated by changing from the variable $\tau$ to the correlation 
length $\xi$ as an independent parameter. Here, for simplicity, we just employ
the $\varepsilon$ expansion. For the extraction of the divergences of the 
vertex function $\Gamma_{1,2}$, the external momenta and frequencies may be set
to zero. The contribution from the diagram in FIG.~\ref{1-loop}(b) then becomes
\begin{equation}
\label{3(b)}
  6(b) = 2 \mathring{\lambda}\mathring{g}^3 \int_\mathbf{p} 
  \frac{1}{\bigl[ 2 (\mathring{\tau} + \mathbf{p}^2) \bigr]^2} =
  \frac{G_\varepsilon}{\varepsilon} \, \mathring{\tau}^{-\varepsilon/2} \,
  \mathring{\lambda} \mathring{g}^{\,3} \, . 
\end{equation}

Combining the zero-loop expressions with the results of this short calculation,
we obtain the one-loop vertex functions to the desired order in $\omega$ and 
$\mathbf{q}^2$:
\begin{align}
\label{G_11}
  \mathring{\Gamma}_{1,1} &= i \omega \Big( 
  1 - \frac{G_\varepsilon}{4 \varepsilon} \, \mathring{g}^2 \, 
  \mathring{\tau}^{-\varepsilon/2} \Big) + \mathring{\lambda} \, \mathbf{q}^2
  \Big( 1 - \frac{G_\varepsilon}{8 \varepsilon} \, \mathring{g}^2 \,
  \mathring{\tau}^{-\varepsilon/2} \Big) \nonumber \\
  &\quad + \mathring{\lambda} \mathring{\tau} \Big( 1 - 
  \frac{G_\varepsilon}{\varepsilon (2-\varepsilon)} \, \mathring{g}^2 \,
  \mathring{\tau}^{-\varepsilon/2} \Big) + \ldots \ ,
\end{align}
and
\begin{equation}
\label{G_12}
  \mathring{\Gamma}_{1,2} = \mathring{\lambda} \, \mathring{g} \Big( 
  1 - \frac{G_\varepsilon}{\varepsilon} \, \mathring{g}^2 \, 
  \mathring{\tau}^{-\varepsilon/2} \Big) + \ldots \, .
\end{equation}
In order to absorb the $\varepsilon$ poles into renormalization factors, we 
employ the scheme (\ref{Ren-Z}), (\ref{Ren-Gamma}) using $\tilde{Z} = Z$. To 
one-loop order we arrive at
\begin{align}
\label{Gamma_11}
  \Gamma_{1,1} &= i\omega \Big[ Z - \frac{u}{4 \varepsilon} \left( 
  \frac{\mu}{\tau} \right)^{\varepsilon/2} \Big] + \lambda \, \mathbf{q}^2
  \Big[ Z_\lambda - \frac{u}{8 \varepsilon} \left( \frac{\mu}{\tau} 
  \right)^{\varepsilon/2} \Big] \nonumber \\
  &\quad + \lambda \, \tau \Big[ Z_\tau - \frac{u}{\varepsilon (2-\varepsilon)}
  \left( \frac{\mu}{\tau} \right)^{\varepsilon/2} \Big] + \ldots \, ,
\end{align}
and
\begin{equation}
\label{Gamma_12}
  G_{\varepsilon} \bigl(\Gamma_{1,2} \bigr)^2 = \mu^{\varepsilon} \lambda^2 \, 
  u \Big[ Z_u - \frac{2u}{\varepsilon} \left( \frac{\mu}{\tau} 
  \right)^{\varepsilon/2} \Big] \, .
\end{equation}
Therefore, the $\varepsilon$ poles are eliminated by the minimal choices
$Z = 1 + u/4 \varepsilon$, $Z_\lambda = 1 + u/8 \varepsilon$, 
$Z_\tau = 1 + u/2 \varepsilon$, and $Z_u = 1 + 2u/\varepsilon$.

\begin{figure}[ptb]
\begin{center}\includegraphics[width=6cm]{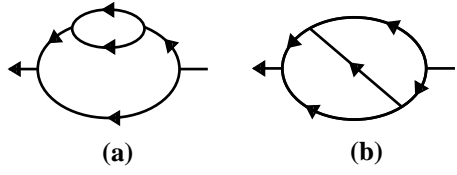}\end{center}
\caption{The two-loop diagrams for the propagator self-energy.}
\label{2-loop1}
\end{figure}

\begin{figure}[b]
\begin{center}\includegraphics[width=11cm]{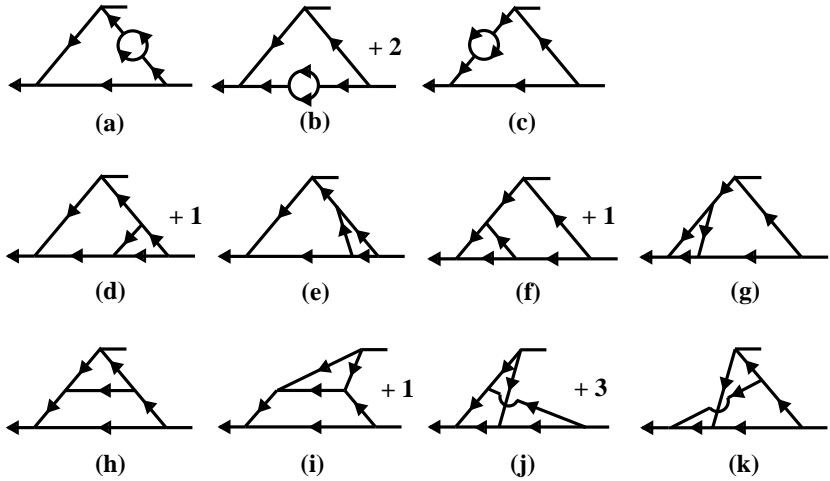}\end{center}
\caption{The two-loop diagrams for the three-point vertex function; the digits 
         indicate the number of different possible time orderings.}
\label{2-loop2}
\end{figure}

The two-loop graphs for the propagator self-energy and the three-point vertex
are depicted in FIG.~(\ref{2-loop1}) and (\ref{2-loop2}), respectively. The
ensuing explicit Z factors read \cite{Ja81,Ja01,BrDa74}
\begin{subequations}
\label{Z-Fakt}
\begin{align}
  Z &= 1 +\frac{u}{4 \varepsilon} + \left( \frac{7}{\varepsilon} - 3 
  + \frac{9}{2} \ln\frac{4}{3} \right) \frac{u^2}{32 \varepsilon} + O(u^3) \, ,
  \label{Z-Fakt_s}\\
  Z_\lambda &= 1 + \frac{u}{8\varepsilon} + \left( \frac{13}{4 \varepsilon} 
  - \frac{31}{16} + \frac{35}{8} \ln\frac{4}{3} \right) 
  \frac{u^2}{32 \varepsilon} + O(u^3) \, , \label{Z-Fakt_l}\\
  Z_\tau &= 1 + \frac{u}{2 \varepsilon} + \left( \frac{16}{\varepsilon} - 5 
  \right) \frac{u^2}{32 \varepsilon} + O(u^3) \, , \label{Z-Fakt_t}\\
  Z_u &= 1 + \frac{2 u}{\varepsilon} + \left( \frac{4}{\varepsilon} - 1 \right)
  \frac{7u^2}{8 \varepsilon} + O(u^3) \, . \label{Z-Fakt_u}
\end{align}
\end{subequations}
The RG functions then follow directly from Eqs.~(\ref{RG-Fu}),
\begin{subequations}
\label{RG-FuDP}
\begin{align}
  \gamma(u) &= -\frac{u}{4} + \left( 2 - 3 \ln\frac{4}{3} \right) 
  \frac{3 u^2}{32} + O(u^3) \, , \label{gamma-DP}\\
  \zeta(u) &= -\frac{u}{8} + \left( 17 - 2 \ln\frac{4}{3} \right)
  \frac{u^2}{256} + O(u^3) \, , \label{zeta-DP}\\
  \kappa(u) &= \frac{3 u}{8} - \left( 7 + 10 \ln\frac{4}{3} \right)
  \frac{7 u^2}{256} + O(u^3) \, , \label{kappa-DP}\\
  \beta(u) &= \Big[ - \varepsilon + \frac{3u}{2} - \left( 169 + 106 
  \ln\frac{4}{3} \right) \frac{u^2}{128} + O(u^3) \Big] u \, . \label{beta-DP}
\end{align}
\end{subequations}

The IR-stable fixed point $u_{\ast}$ is determined via $\beta(u_\ast) = 0$ by 
means of the $\varepsilon$ expansion,
\begin{equation}
\label{fixpt}
  u_\ast = \frac{2 \varepsilon}{3} \Big[ 1 + \left( \frac{169}{288} + 
  \frac{53}{144}\ln\frac{4}{3} \right) \varepsilon + O(\varepsilon^2) \Big]\, .
\end{equation}
As a consequence, the general asymptotic scaling results (\ref{FP-Werte}) and
(\ref{As-Skal}) are finally found to carry the $\varepsilon$-expanded special 
DP critical exponents
\begin{subequations}
\label{DP-Exp}
\begin{align}
  \eta &= - \frac{\varepsilon}{6} \Big[ 1 + \left( \frac{25}{288} +
  \frac{161}{144} \ln\frac{4}{3} \right) \varepsilon +O(\varepsilon^2) \Big] \,
  , \label{eta-DP}\\
  z &= 2 - \frac{\varepsilon}{12} \Big[ 1 + \left( \frac{67}{288} +
  \frac{59}{144} \ln\frac{4}{3} \right) \varepsilon + O(\varepsilon^2) \Big] \,
  , \label{z-DP}\\
  \nu &= \frac{1}{2} + \frac{\varepsilon}{16} \Big[ 1 + \left( \frac{107}{288}
  - \frac{17}{144} \ln\frac{4}{3} \right) \varepsilon + O(\varepsilon^2) \Big]
  \, , \label{nu-DP}\\
  \beta &= \nu \, \frac{d+\eta}{2} = 1 - \frac{\varepsilon}{6} \Big[ 1 - \left(
  \frac{11}{288} - \frac{53}{144} \ln\frac{4}{3} \right) \varepsilon + 
  O(\varepsilon^2) \Big] \, ,
\label{bet-DP}
\end{align}
\end{subequations}
with $\delta_{N,\tilde{N}} = (N + \tilde{N}) \beta/\nu$.

\subsection{Critical Properties of Directed Percolation}

Equipped with the important general asymptotic scaling results for the DP 
Green's functions, which followed directly from the knowledge of the Z factors,
we are now in the position to determine the critical properties of dynamic as 
well as static observables.

\subsubsection{Dynamic Observables}

Using the asymptotic scaling form (\ref{As-Skal}), the number $N(t,\tau)$ of
\emph{active particles} generated by a seed at the origin is given by
\begin{align}
\label{N(t)}
  N(t,\tau) &= \int \! d^dr \, G_{1,1}(\mathbf{r},t;\tau,u;\lambda,\mu)
  \nonumber\\
  &= \mu^{-d} A \, \tilde{A} \, l^{2\beta/\nu-d} 
  \tilde{F}_{1,1}(\mathbf{q}=0, l^{z}A_{\lambda} \, \lambda\mu^2 \, t,
  l^{-1/\nu} A_{\tau} \, \tau/\mu^2) \nonumber\\
  &= A_N \, t^{\theta_s} \, \Phi_N(B_\tau \tau \, t^{1/\nu_{\parallel}}) \, .
\end{align}
Here we have introduced the exponents
\begin{equation}
\label{SkalExp-1}
  \nu_\parallel = z \, \nu \, , \qquad \theta_s = \gamma^{\prime} / z \nu \, ,
  \qquad \gamma^{\prime} = d \nu - 2 \beta \, , 
\end{equation}
and the non-universal amplitude combinations
\begin{equation}
\label{Ampl-1}
  A_t = \lambda \mu^2 A_\lambda \, , \quad A_N = \mu^{-d} A_t^{\theta_s} A \,
  \tilde{A} \, \tilde{F}_{1,1}(0,1,0) \, , \quad B_\tau = \mu^{-2} 
  A_t^{1/\nu_\parallel} A_\tau \, .
\end{equation}
$\Phi_{N}(x)$ is a universal scaling function normalized to $\Phi_N(0)=1$.

As shown in Ref.~\cite{Ja03}, the \emph{survival probability} $P(t,\tau)$ of an
active cluster emanating from a seed at the origin is asymptotically given by
\begin{equation}
\label{Surv.P.}
  P(t,\tau) = -\lim_{k \to \infty} \langle\mathrm{e}^{-k \mathcal{N}}
  \tilde{s}(\mathbf{r}=0,-t) \rangle = - G_{0,1}(\mathbf{0},\mathbf{-}t;\tau,
  k=\infty,u;\lambda,\mu) \, , 
\end{equation}
with $\mathcal{N} = \int \! d^dx \, s(\mathbf{x},0)$, and where the Green's 
function $G_{0,1}$ is calculated with the response functional 
$\mathcal{J}_k = \mathcal{J}_{\rm DP} + k \mathcal{N}$. Note that because of 
the duality invariance (\ref{ZeitSp}) the survival probability $P(t,\tau)$ is 
in fact related to the mean asymptotic density of the active particles 
$\rho(t,\tau;\rho_0)$ for a process which starts with a homogeneous active 
density $\rho_0$ by
\begin{equation}
\label{P-rho}
  A \, P(t,\tau) = \tilde{A} \, \rho(t,\tau;\infty) \, .
\end{equation}
Recalling the scaling form (\ref{As-Skal}) of $G_{0,1}$ we deduce that
\begin{equation}
\label{P(t)}
  P(t,\tau) = A_P \, t^{-\delta_s} \, \Phi_P(B_\tau \, \tau \, 
  t^{1/\nu_\parallel}) \, ,
\end{equation}
with
\begin{equation}
\label{Skal-Ampl-2}
  \delta_s = \beta / z \nu \, , \qquad A_P = A_t^{-\delta_s} \tilde{A} \,
  F_{0,1}(0,1,0,\infty) \, .
\end{equation}
Again, $\Phi_P(x)$ represents a universal function normalized to 
$\Phi_P(0)=1$. $\Phi_P(x)$ tends to zero exponentially for $x \to \infty$, and 
asymptotically to $C_P |x\|^\beta$ for $x \to -\infty$, where $C_P$ is a 
universal constant. Thus, we find the percolation probability, for $\tau < 0$,
\begin{equation}
\label{PerkProb}
  P_\infty(\tau) = A_\infty |\tau|^\beta \, , \qquad
  A_\infty = A_P B_\tau^\beta C_P \, .
\end{equation}

The extension of a active cluster at time $t$ generated by a seed at the origin
is measured by the \emph{radius of gyration} $R(t,\tau)$ of the active 
particles, as defined via
\begin{equation}
\label{R-def}
  R^2(t,\tau) = \frac{\int \! d^dr \, \mathbf{r}^2 \, G_{1,1}(\mathbf{r},t)}
  {2d \int \! d^dr \, G_{1,1}(\mathbf{r},t)} = 
  - \frac{\partial \ln \tilde{G}_{1,1}(\mathbf{q},t)}{\partial q^2} 
  \bigg\vert_{\mathbf{q}=0} \, .
\end{equation}
From the asymptotic scaling law (\ref{As-Skal}) we infer
\begin{align}
  \frac{\partial \ln \tilde{G}_{1,1}(\mathbf{q},t)}{\partial q^2} 
  \bigg\vert_{\mathbf{q}=0} &= (l\mu)^{-2} \, \frac{\partial \ln 
  \tilde{F}_{1,1}(\mathbf{q},l^z A_{\lambda} \, \lambda \mu^2 \, t, l^{-1/\nu}
  A_\tau \, \tau/\mu^2)}{\partial q^2} \bigg\vert_{\mathbf{q}=0} \nonumber\\
  &= (l\mu)^{-2} \, f_R(l^z A_{\lambda} \, \lambda \mu^2 \, t, l^{-1/\nu}
  A_\tau \, \tau/\mu^2) \, ,
\end{align}
whence for the radius of gyration we obtain the asymptotic scaling form
\begin{equation}
\label{R(t)}
  R^2(t,\tau) = A_R \, t^{z_s} \, \Phi_R(B_\tau \, \tau \, t^{1/\nu_\parallel})
  \, ,
\end{equation}
with
\begin{equation}
\label{Skal-Ampl-3}
  z_s = 2 / z \, , \qquad A_R = \mu^{-2} A_t^{z_s} f_R(1,0) \, ,
\end{equation}
where again $\Phi_R(x)$ is a universal function normalized to $\Phi_R(0)=1$.
The four quantities $A_N$, $A_P$, $A_R$, and $B_\tau$ define a measurable 
complete set of non-universal amplitudes for any system belonging to the DP 
universality class. Once these amplitudes are determined, all other observables
attain universal values.

The last dynamic observable we consider here is the \emph{active density}
$\rho(t,\tau;\rho_0)$ for finite initial $\rho_0$,
\begin{equation}
\label{in-dens}
  \rho(t,\tau;\rho_0) = G_{1,0}(\mathbf{0},t;\tau,\rho_0,u;\lambda,\mu) \, .
\end{equation}
The initial density $\rho_0$ is introduced into the response functional
(\ref{JDP}) via a source $h(\mathbf{r},t) = \lambda^{-1} \rho_0 \, \delta(t)$. 
No new initial renormalization \cite{JSS89} is involved since the perturbation
theory is based solely on the causal propagator (the correlators do not enter),
and initial correlations are irrelevant. Hence, according to the scheme
(\ref{Ren-Z}), the renormalization of $\rho_0$ is given by
\begin{equation}
\label{ren-in-dens}
  \mathring{\rho}_0 = \tilde{Z}^{-1/2} \rho_0 \, ,  
\end{equation}
which leads to an additional derivative term $\frac{1}{2} \tilde{\gamma} \,
\rho_0 \, \partial/\partial\rho_0$ in the RGE (\ref{RGE}). Thus a new 
dependence on $\tilde{X}(l)^{1/2} \rho_0$ and 
$\tilde{A} \, l^{(\tilde{\eta}-d)/2}\rho_{0}/\mu^{d/2}$ arises in 
Eqs.~(\ref{RG-Sol}) and (\ref{As-Skal}), respectively, which leads to
\begin{equation}
\label{rho}
  \rho(t,\tau;\rho_0) = A_\rho \, t^{-\delta_s} \, \Phi_\rho(B_\tau \tau \,
  t^{1/\nu_\parallel},B_\rho \, \rho_0 \, t^{\delta_s + \theta_i}) \, ,
\end{equation}
where $\Phi_\rho(x,y)$ is universal with $\Phi_\rho(x,\infty) = \Phi_P(x)$
and $\Phi_\rho(x,y) = \Phi_\rho^\prime(x) \, y + O(y^2)$. The non-universal
amplitudes are
\begin{equation}
\label{Skal-Ampl-4}
  A_\rho = (A / \tilde{A}) \, A_P \, , \qquad B_\rho = \mu^{-d/2} \tilde{A} \,
  A_t^{\delta_s + \theta_i} \, , 
\end{equation}
and the initial scaling exponent is
\begin{equation}
\label{init-exp}
  \theta_i = - \eta / z \, . 
\end{equation}

\subsubsection{Static Observables}

A steady state with DP dynamics can be generated by introducing a homogeneous
and time-independent external source of activity $h$ in the response functional
(\ref{JDP}). In such a steady state one can then measure static, 
time-independent observables, such as the mean density of activated particles 
and their fluctuations \cite{LuWi02}. Recall, however, that DP defines genuine
non-equilibrium systems; hence, there is no fluctuation-dissipation theorem 
that would relate the correlations of the order parameter to its response to an
external conjugated field. Dynamic correlation and response functions therefore
constitute independent observables.

In order to obtain the \emph{equation of state}, i.e., the order parameter $M$ 
as a function of $\tau$ and $h$ in the steady state, we perform the variable 
shift $s \to M+s$ and determine $M = G_{1,0}(\tau,h,u,\lambda,\mu)$ by the 
`no-tadpole' requirement $\langle s \rangle = 0$. After this shift, the linear
and harmonic part of the response functional become
\begin{align}
\label{Shift-JDP}
  \mathcal{J}_{{\rm DP},0} = \int \! d^dr \, dt \, \Big\{ &\tilde{s} \Big[
  \partial_t + \lambda( \tau + g M - \nabla^2) \Big) s - \frac{\lambda gM}{2}
  \, \tilde{s}^2 \nonumber\\ 
  &\quad + \lambda \Big[ M \Bigl( \tau + \frac{gM}{2} \Bigr) - h \Big] 
  \tilde{s} \Big\} \, .
\end{align}
As a consequence, aside from the propagator the perturbation expansion is now
based on a correlator induced by the noise vertex $\sim \tilde{s}^2$. We will
present no details of the calculation here, but merely state the equation of 
state in a parametric form, as derived in a two-loop approximation in 
Ref.~\cite{JKO98}:
\begin{equation}
\label{Eq.ofSt.}
  \tau = a R \, (1-\theta) \, , \quad M = R^\beta \theta \, , \quad
  h = b R^\Delta \, \theta(2-\theta) + O(\varepsilon^3) \, ,
\end{equation}
where $a$ and $b$ are non-universal amplitudes that can be related to the
previously introduced fundamental four amplitudes, and the exponent $\Delta$ is
given by
\begin{equation}
\label{Delta+Gamma}
  \Delta = \beta + \gamma \, , \qquad \gamma = (z-\eta) \nu \, . 
\end{equation}
The parameter $R \geq 0$, which measures the distance to the critical point, 
and the parameter range $0 \leq \theta \leq 2$ are required to describe the 
entire phase diagram in the critical region.

As a simple application of this parametric representation we briefly discuss
the susceptibility $\chi = \partial M / \partial h |_\tau$,
which satisfies a power law for $h \to 0$,
\begin{equation}
\label{as-suscept}
  \chi = \chi_\pm |\tau|^{-\gamma} \, , 
\end{equation}
with amplitudes $\chi_+$ and $\chi_-$ that correspond to the cases $\tau>0$ and
$\tau<0$, respectively. We obtain
\begin{equation}
\label{sus_expl}
  b \, \chi = R^{-\gamma} \, \frac{1-(1-\beta) \, \theta}
  {\Delta \, \theta (2-\theta) + 2(1-\theta)^2} \, . 
\end{equation}
Therefore the universal amplitude ratio $\chi_- / \chi_+$ can be expressed
to order $\varepsilon^2$ in terms of the  order parameter exponent $\beta$ as
\begin{equation}
\label{Ratio}
  \frac{\chi_-}{\chi_+} = 2 \beta - 1 + O(\varepsilon^3) = 
  1 - \frac{\varepsilon}{3} \bigl[ 1 + 0.067 \, \varepsilon + O(\varepsilon^2) 
  \bigr] \, . 
\end{equation}

The susceptibility $\chi$ may also be represented by the integral over space 
and time of the Green's function $G_{1,1}$:
\begin{equation}
\label{susc-int}
  \chi = \lambda \int\! d^dr \,dt \, G_{1,1}(\mathbf{r},t;\tau,h,u,\lambda,\mu)
  = |\tau|^{-\gamma} f_\pm(|\tau|^{-\Delta} h) \, . 
\end{equation}
In contrast, the mean-square fluctuation 
$\chi^{\prime} = \langle (\Delta N)^2 \rangle / V$ of the activated particle
number are given by
\begin{equation}
\label{fluct-int}
  \chi^{\prime} = \int \! d^dr \, dt \, G_{2,0}(\mathbf{r},t;\tau,h,u,\lambda,
  \mu) = |\tau|^{-\gamma^{\prime}} f_\pm^{\prime}(|\tau|^{-\Delta} h) \, ,
\end{equation}
where the exponent $\gamma^{\prime}$ is defined by Eq.~(\ref{SkalExp-1}) and
differs from $\gamma$. A one-loop calculation yields
\begin{equation}
\label{fluctuat}
  b^{\prime} \, \chi^{\prime} = R^{-\gamma^{\prime}} \,
  \frac{\theta}{1 + \alpha \, \theta} \, ,
\end{equation}
with the non-universal parameter $b^{\prime}$ and $\alpha = \bigl( \frac{1}{6} 
+ \frac{1}{4} \, \ln3/2 \bigr) \, \varepsilon + O(\varepsilon^2)$.

\subsubsection{Logarithmic Corrections at the Upper Critical Dimension}

Above the upper critical dimension $d_c = 4$, where mean-field theory is
applicable, the coupling constant $g$ tends to zero under the renormalization
group transformation. However, $g$ represents a \emph{dangerously irrelevant 
variable} here, since it scales various observables, and setting $g = 0$ 
rigorously leads either to zero or infinity for relevant quantities. The 
twofold nature of $g$ as both a relevant scaling variable and an irrelevant 
loop-expansion generating parameter is lucidly exposed by writing the 
generating functional for the vertex functions (\ref{Vert-Kum}) in the form
\begin{equation}
\label{Phi}
  \Gamma[\tilde{s},s;\tau,g] = g^{-2} \, \Phi[g \tilde{s},g s;\tau,u] \, .
\end{equation}
The expansion of the functional $\Phi\lbrack\tilde{\varphi},\varphi;u]$ into a
series with respect to $u = G_{\varepsilon} \mu^{-\varepsilon} g^2$ yields the
loop expansion. The zeroth-order term $g^{-2} \Phi[g \tilde{s},g s;0]$ is just
the response functional (\ref{J3}) itself, i.e., $\mathcal{J}$ represents the 
mean-field contribution to the dynamic `free energy' $\Gamma$. The scaling form
of the generating functional (\ref{KumGen}) for the cumulants that corresponds
to (\ref{Phi}) reads
\begin{equation}
\label{Omega}
  \mathcal{W}[H,\tilde{H};\tau,g] = g^{-2}  \Omega[g H,g \tilde{H};\tau,u] \, .
\end{equation}
To leading order in the logarithmic corrections, we may neglect the dependence 
of $\Omega$ and $\Phi$ on $u$. (For the next-to-leading corrections see 
Ref.~\cite{JaSt04}.)

Solving the characteristic equation (\ref{Char}) for $\bar{u}$ at the upper
critical dimension ($\varepsilon = 0$) yields to leading order for $l \to 0$:
\begin{equation}
\label{u-as}
  \bar{u}(l) = 1 / \beta_2 |\ln l| \, , 
\end{equation}
where we introduce the Taylor expansion $f(u) = f_0 + f_1 u + f_2 u^2 +\ldots$ 
for any of the RG flow functions $f = \gamma, \zeta, \kappa, \beta$. Thus, from
Eq.~(\ref{beta-DP}), $\beta_2 = 3/2$ for DP. The remaining characteristics are 
all of the same structure, namely
\begin{equation}
\label{Char4}
  l \frac{d\ln Q}{dl} = q(\bar{u}) \, . 
\end{equation}
Here, $Q$ stands for $X$, $\ln \bar{\tau}$, and $\ln\bar{\lambda}$, 
respectively, whereas $q$ represents either $\gamma$, $\kappa$, or $\zeta$, as
defined in Eq.~(\ref{RG-Fu}). To leading order, Eq.~(\ref{Char4}) is solved by
\begin{equation}
\label{Q(l)}
  Q(l) \propto |\ln l|^{-q_1 / \beta_2} Q(1) \, . 
\end{equation}
Combining everything we obtain asymptotically
\begin{align}
\label{W-as}
  &\mathcal{W}[H,\tilde{H};\tau,g;\lambda,\mu] \simeq |\ln l| \nonumber \\
  &\qquad \times \Omega\big[ |\ln l|^{-5/12} H, |\ln l|^{-5/12} \tilde{H}; 
  |\ln l|^{-1/4} \tau,0; |\ln l|^{1/12} \lambda, l \mu \big] \, , 
\end{align}
where the non-universal amplitudes have been absorbed into the variables. 
Taking the required functional derivatives at $\tilde{H} = 0$, 
$H(\mathbf{r},t) = \lambda h + \rho_0 \, \delta(t)$ and employing dimensional
scaling then yields the asymptotic Green's functions with the logarithmic
scaling corrections at $d_c = 4$:
\begin{align}
\label{log-Gr}
  &G_{N,\tilde{N}}(\{\mathbf{r},t\},\tau,u,\lambda,\mu) \propto |\ln l| 
  (l^2 |\ln l|^{-5/12})^{N+\tilde{N}} \nonumber \\ &\times 
  F_{N,\tilde{N}}(\{ l\mathbf{r}, l^2 |\ln l|^{1/12} t\}; l^{-2} |\ln l|^{-1/4}
  \tau, l^{-4} |\ln l|^{-1/2} h, l^{-2} |\ln l|^{-5/12} \rho_0) \, .
\end{align}

From this general result we find for the dynamical observables
\begin{subequations}
\label{log-korr-dyn}
\begin{align}
  N(t,\tau) &= (\ln t)^{1/6} f_N \big( (\ln t)^{-1/6} \, t \tau \big) \, ,
  \label{log-korr-N}\\
  P(t,\tau) &= \frac{(\ln t)^{1/2}}{t} \, f_P\big( (\ln t)^{-1/6} \, t \tau
  \big) \, , \label{log-korr-P}\\
  R^2(t,\tau) &= t \, (\ln t)^{1/12} \, f_P\big( (\ln t)^{-1/6} \, t \tau \big)
  \, , \label{log-korr-R}\\
  \rho(t,\tau) &= \frac{(\ln t)^{1/2}}{t} \, f_\rho\big( (\ln t)^{-1/6} \, 
  t \tau, (\ln t)^{-1/3} \,t \rho_0) \nonumber\\
  &= (\ln t)^{1/6} \, \rho_0 \, \bar{f}_\rho\big( (\ln t)^{-1/6} \, t\tau,
  (\ln t)^{-1/3} \, t \rho_0 \big) \label{log-korr-rho} \, .
\end{align}
\end{subequations}
The logarithmic correction for the percolation probability becomes
\begin{equation}
\label{log-korr-Perk}
  P_\infty(\tau) = C \theta(-\tau) \, |\tau| |\ln |\tau||^{1/3} \, . 
\end{equation}
and the equation of state and the fluctuations in the steady state read to
leading order
\begin{subequations}
\label{log-korr-stat}
\begin{align}
  M &= A \bigl[ \sqrt{\tau^2 + B h} - \tau \bigr] \, 
  \big\vert \ln\sqrt{\tau^2 + B h} \big\vert^{1/3} \, , \label{log-korr -EQS}\\
  \chi^\prime &= A^\prime \bigg[ 1 - \frac{\tau}{\sqrt{\tau^2 + B h}} \bigg] \,
  \big\vert \ln\sqrt{\tau^2 + B h} \big\vert^{1/6} \, , \label{log-korr-fluc}
\end{align}
\end{subequations}
where $A$, $A^\prime$, $B$, and $C$ are non-universal amplitudes.

\subsubsection{Finite-Size Scaling}

We end this section by discussing a method that allows the explicit calculation
of \emph{finite-size effects} in DP \cite{JSS88}. We will consider the model 
defined by the action (\ref{JDP}), in a finite cubic geometry of linear size 
$L$ with periodic boundary conditions. Expanding the fields in Fourier modes
\begin{equation}
 \label{FouMod}
  s(\mathbf{r},t) = \sum_{\mathbf{q}} \mathrm{e}^{i \mathbf{q}\cdot\mathbf{r}}
  s(\mathbf{q},t) \, , \qquad \tilde{s}(\mathbf{r},t) = \sum_{\mathbf{q}}
  \mathrm{e}^{i \mathbf{q}\cdot\mathbf{r}} \tilde{s}(\mathbf{q},t) \, ,
\end{equation}
where each component of $\mathbf{q}$ only assumes discrete values, namely
multiples of $2\pi / L$, it is clear that the $(\mathbf{q}=0)$-mode cannot be
treated perturbatively at the critical point $\tau = h =0$, since the
propagator displays an isolated pole at $\mathbf{q} = 0$. Therefore, in order 
to evaluate finite-size effects, one has to construct an effective dynamic 
response functional for the $(\mathbf{q}=0)$-mode (which subsequently has to be
treated non-perturbatively) by tracing out all modes with $\mathbf{q} \neq 0$.

Upon decomposing the fields
\begin{subequations}
\label{Sep}
\begin{align}
  s(\mathbf{r},t) = \Phi(t) + \varphi(\mathbf{r},t) \, , &\qquad 
  \tilde{s}(\mathbf{r},t) =\tilde{\Phi}(t) + \tilde{\varphi}(\mathbf{r},t) \, ,
  \label{Sep1}\\
  \text{with} \ \int \! d^dr \,\varphi(\mathbf{r},t) &= \int \! d^dr \, 
  \tilde{\varphi}(\mathbf{r},t) = 0 \, , \label{Sep2}
\end{align}
\end{subequations}
one can separate the $(\mathbf{q}=0)$-modes $\Phi(t)$ and $\tilde{\Phi}(t)$ 
from their orthogonal complements $\varphi(\mathbf{r},t)$ and 
$\tilde{\varphi}(\mathbf{r},t)$. After performing this decomposition in the 
action $\mathcal{J}_{\rm DP}$ we obtain
\begin{subequations}
\label{Decomp}
\begin{align}
  \mathcal{J}_{\rm DP}^{(0)} &= L^d \int \! dt \, \biggl\{ \tilde{\Phi} \bigg[
  \partial_t + \lambda \, \tau + \frac{\lambda g}{2} \bigl( \Phi - \tilde{\Phi}
  \bigr) \bigg] \Phi - \lambda \, h \, \tilde{\Phi} \biggr\} \, , \label{J-0}\\
  \mathcal{J}_{\rm DP}^{(1)} &= \int \! dt \, d^dr \,\biggl\{ \tilde{\varphi}
  \bigg[ \partial_t + \lambda (\tau - \nabla^2) + \frac{\lambda g}{2} (\Phi -
  \tilde{\Phi}) \bigg] \varphi + \frac{\lambda g}{2} \bigl( \tilde{\Phi}
  \varphi^2 - \Phi \tilde{\varphi}^2 \bigr) \biggr\} \, , \label{J-1}\\
  \mathcal{J}_{\rm DP}^{(2)} &= \frac{\lambda g}{2} \int \! dt \, d^dr \,
  \tilde{\varphi} \bigl( \varphi - \tilde{\varphi} \bigr) \varphi \, . 
  \label{J-2}
\end{align}
\end{subequations}
Integrating over all modes with $\mathbf{q} \neq 0$ we arrive at the 
\emph{effective action}
\begin{equation}
\label{J-eff}
  \mathcal{J}_{\rm DP}^{({\rm eff})} = - \ln \int \! \mathcal{D}
  [\tilde{\varphi},\varphi] \exp\bigl(-\mathcal{J}_{\rm DP}^{(0)}-
  \mathcal{J}_{DP}^{(1)}-\mathcal{J}_{\rm DP}^{(2)} \bigr) = 
  \mathcal{J}_{\rm DP}^{(0)}[\tilde{\Phi},\Phi] + \Sigma[\tilde{\Phi},\Phi]\, .
\end{equation}
The contribution $\Sigma[\tilde{\Phi},\Phi]$ can now be analyzed perturbatively
by means of a double expansion in powers of the fields $\tilde{\Phi}$ and 
$\Phi$ that arise in the vertices of $\mathcal{J}_{\rm DP}^{(1)}$, and in the
number of loops due to insertion of extra vertices originating from
$\mathcal{J}_{\rm DP}^{(2)}$. Up to and including terms of third order in 
$\tilde{\Phi}$, $\Phi$, and after renormalization, one finds \cite{JSS88}
\begin{equation}
\label{J-eff1}
  \mathcal{J}_{\rm DP}^{({\rm eff})} = L^d \int \! dt \, \biggl\{ \tilde{\Phi}
  \Big[ \hat{r} \, \partial_t + \lambda \, \hat{\tau} + 
  \frac{\lambda \hat{g}}{2} \bigl( \Phi - \tilde{\Phi} \bigr) \Big] \Phi
  - \lambda \, h \, \tilde{\Phi} \biggr\} \, . 
\end{equation}
In dimensions $d \leq 4$, the functions $\hat{r}(\tau,L)$, 
$\hat{\tau}(\tau,L)$, and $\hat{g}(\tau,L)$ display scaling properties that 
follows from the RGE. For $d > 4$, the contributions stemming from 
$\Sigma[\tilde{\Phi},\Phi]$ can be neglected asymptotically, whence 
$\hat{r} \simeq 1$, $\hat{\tau} \simeq \tau$, and $\hat{g} \simeq g$.

We proceed by rescaling the fields and the time scale according to
\begin{subequations}
\label{FP-Skal}
\begin{align}
  \Phi(t) &= \alpha \, M(s) \, , \qquad 
  \tilde{\Phi}(t) = \alpha \, \tilde{M}(s) \, , \qquad 
  \lambda t = \beta s \, , \label{FP-Skal1}\\
  \alpha &= \hat{r}^{-1/2} L^{-d/2} \, , \qquad \beta = 2 \hat{g}^{-1} \,
  \hat{r}^{3/2} L^{d/2} \, , \label{FP-Skal2}
\end{align}
\end{subequations}
and thereby obtain for the effective response functional
\begin{equation}
\label{Funk-FP}
  \mathcal{J}_{\rm DP}^{({\rm eff})} = \int \! ds \, \biggl\{ \tilde{M} \Big[
  \partial_s + a + \bigl( M - \tilde{M} \bigr) \Big] M - b \, \tilde{\Phi}
  \biggr\} \, ,
\end{equation}
where the two parameters $a$ and $b$ are given by
\begin{subequations}
\label{FP-Parame}
\begin{align}
  a &= 2 \hat{r}^{-1/2} \, \hat{g}^{-1} \, \hat{\tau} \, L^{d/2} =
  a_0 + a_1 \tau L^{1/\nu} + O\big( (\tau L^{1/\nu})^2 \big) \, ,
  \label{FP-Param1}\\
  b &= 2 \hat{r} \, \hat{g}^{-1} \, h \, L^d = \big( b_0 + O(\tau L^{1/\nu}) 
  \big) h \, L^{\Delta/\nu} \, , \label{FP-Param2}
\end{align}
\end{subequations}
and
\begin{equation}
\label{FP-alpha}
  \alpha = \hat{r}^{-1/2} L^{-d/2} = \big( \alpha_0 + O(\tau L^{1/\nu}) \big)
  L^{-\beta/\nu} \, .
\end{equation}
For $d > 4$, we have $a \propto g^{-1} L^{d/2} \, \tau + c g \, L^{2-d/2}$ and
$b \propto g^{-1} L^d \, h$. Hence, besides the finite-size scaling of the 
control parameter $\tau$ and the source $h$, an additional shift $a_0$ results 
from the elimination of the $(\mathbf{q}\neq0)$-modes. However, above four 
dimension this shift is proportional to $gL^{2-d/2}$ and represents the leading
correction. The Green's functions for the spatially averaged variables 
$\Phi(t)$ and $\tilde{\Phi}(t)$ are now calculated with the response functional
$\mathcal{J}_{\rm DP}^{({\rm eff})}[\tilde{M},M;a,b]$, Eq.~(\ref{Funk-FP}). 
Consequently
\begin{equation}
\label{Kum}
  G_{N,\tilde{N}}(\{ t \},\tau,h,g,L) = \langle [\Phi]^N 
  [\tilde{\Phi}]^{\tilde{N}} \rangle^{({\rm cum})} = \alpha^{N+\tilde{N}} 
  F_{N,\tilde{N}}(\{ 2s \},a,b) \, ,
\end{equation}
and above four dimension, i.e., in the mean-field region, we obtain 
asymptotically
\begin{equation}
\label{Kum-mf}
  G_{N,\tilde{N}}(\{ t \},\tau,h,g,L) = L^{-(N+\tilde{N}) d/2} 
  F_{N,\tilde{N}}(\{ \lambda g L^{-d/2} \, t \},g^{-1} L^{d/2} \, \tau, 
  2g^{-1} L^d \, h) \, . 
\end{equation}

The scaling functions $F_{N,\tilde{N}}$ can be evaluated by an alternative
approach. Via the equivalence of the response functional (\ref{Funk-FP}) with 
the It\^o-Langevin equation, which in turn is equivalent to a corresponding 
Fokker-Planck equation for the probability density $\mathcal{P}(M,s)$ of the 
stochastic variable $M$, we obtain
\begin{equation}
\label{FokPl}
  \frac{\partial}{\partial s} \mathcal{P}(M,s) = \frac{\partial}{\partial M}
  \Big\{ \big[ (a+M) M - b \big] \mathcal{P}(M,s) \Big\} + 
  \frac{\partial^2}{\partial M^2} \big[ M \mathcal{P}(M,s) \big] \, . 
\end{equation}
In particular, the stationary solution $\mathcal{P}_{\rm st}(M)$ for $b > 0$ is
easily found,
\begin{equation}
\label{Prob-stat}
   \mathcal{P}_{\rm st}(M) = C M^{b-1} \exp\bigg[ - \Bigl( a +\frac{1}{2} M
   \Bigr) M \bigg] \, , 
\end{equation}
where $C$ represents a finite normalization constant. All the moments of the 
averaged agent density $\rho$ can now be expressed in terms of parabolic 
cylinder functions. In particular, for $d > 4$, and at criticality ($a = 0$), 
using $\rho \propto \Phi \propto L^{-d/2} M$, one arrives at
\begin{equation}
\label{Momente}
  \langle \rho^N \rangle = \big( A_\rho \, L^{-d/2} \big)^N \,
  \frac{\Gamma(N/2 + A_h \, L^d h)}{\Gamma(A_h \, L^d h)} \, ,
\end{equation}
where $A_\rho$ and $A_h$ are non-universal amplitudes, and $\Gamma(x)$ denotes 
Euler's Gamma function. Specifically, Binder's cumulant $Q$ is given by  the 
simple expression
\begin{equation}
\label{Binder-Kum}
  Q:= 1 - \frac{\langle \rho^4 \rangle}{3 \langle \rho^2 \rangle^2} =
  \frac{2}{3} - \frac{1}{A_h \, L^d h} \, . 
\end{equation}
This function has been measured to high precision in recent simulations 
\cite{Lu04}.

\section{Field Theory of Dynamic Isotropic Percolation}

We consider now the field theory of dynamic isotropic percolation. We are
interested in the dynamic as well as static (emerging as $t \to \infty$) 
properties of isotropic percolation. As mentioned earlier, the static debris
clusters after the activation process have ceased are described by the usual
static percolation theory near the critical threshold. Of course, we could base
our entire RG analysis on the full dynamic functional $\mathcal{J}_{\rm dIP}$ 
as given in Eq.~(\ref{JdIP}), and extract the static from the full dynamic 
behavior by taking the limit $t \to \infty$ in the end. This would imply, 
however, that we would have to determine all the required renormalizations from
the dynamic Feynman graphs, composed of the diagrammatic elements encoded in
the action $\mathcal{J}_{\rm dIP}$. Fortunately, there is a considerably more 
economic approach which is based on taking the so-called \emph{quasi-static 
limit}. Imagine that we initiate a process through external activity sources 
$\lambda \, h(\mathbf{r},t) = k(\mathbf{r}) \, \delta(t)$, localized in time at
$t = 0$; but we are interested only at Green's functions of the debris at 
$t = \infty$. We will see shortly that the perturbation expansion simplifies 
tremendously in this limit. All renormalization factors but one can be 
calculated directly using this much simpler method. Thus, only for the single 
remaining renormalization do we have to resort to the full dynamic response 
functional $\mathcal{J}_{\rm dIP}$. Taking the quasi-static limit amounts to 
switching the fundamental field variable from the agent density to the final 
density of debris $\varphi(\mathbf{r}) := S(\mathbf{r},\infty) = \lambda 
\int_0^\infty dt \, s(\mathbf{r},t)$ that is ultimately left behind by the 
epidemic, and the associated response field 
$\tilde{\varphi}(\mathbf{r}) = \tilde{s}(\mathbf{r},0)$. In particular, the 
Green's functions corresponding to the correlation functions
$\langle \prod_i \varphi(\mathbf{r}_i) \tilde{\varphi}(\mathbf{0}) \rangle$ 
will turn out to be important for our analysis, since they actually encode the 
static properties of the debris percolation cluster emanating from a seed that
is localized at the origin at time $t = 0$.

\subsection{Quasistatic Field Theory}

\subsubsection{Quasistatic Hamiltonian}

Following up on the previous remarks, we now proceed to formally take the 
quasi-static limit of the dynamic functional for dIP. The structure of 
$\mathcal{J}_{dIP}$ allows us to directly let
\begin{equation}
\label{quaslim}
  \tilde{s}(\mathbf{r},t) \to \tilde{\varphi}(\mathbf{r}) \, , \qquad
  \varphi(\mathbf{r}) = \lambda\int_{0}^{\infty}dt\,s(\mathbf{r},t)\,.
\end{equation}
This procedure leads us from the action (\ref{JdIP}) directly to the 
\emph{quasi-static Hamiltonian} with source $k$
\begin{equation}
\label{Hamilt}
  \mathcal{H}_{\rm IP} = \int \! d^dx \, \Big\{ \tilde{\varphi} \Big[
  \tau - \nabla^2 + \frac{g}{2} \bigl( \varphi - \tilde{\varphi} \bigr) \Big]
  \varphi - \, k \tilde{\varphi} \Big\} \, .
\end{equation}
By just considering the involved time integrations from `left' (i.e., the 
largest time involved) to `right' (the smallest time) in any graph, it is easy
to see that $\mathcal{H}_{\rm IP}$ in fact generates all diagrams contributing 
to $\langle \prod_i S(\mathbf{r}_i,\infty) \prod_j 
\tilde{s}(\mathbf{\tilde{r}}_j,0) \rangle = \langle \prod_i 
\varphi(\mathbf{r}_i) \prod_j \tilde{\varphi}(\mathbf{\tilde{r}}_j) \rangle$. 
By itself, however, the Hamiltonian (\ref{Hamilt}) is insufficient to describe 
the static properties of isotropic percolation (IP). As a remnant of its 
dynamical origin, $\mathcal{H}_{\rm IP}$ must be supplemented with the causality rule 
that forbids closed propagator loops.

The propagator of the quasi-static theory immediately follows from the
Hamiltonian,
\begin{equation}
\label{statProp}
  \langle \varphi(\mathbf{r}) \tilde{\varphi}(\mathbf{r}^\prime) \rangle_0 = 
  \int_{\mathbf{q}} \frac{\exp \bigl( i \mathbf{ q \cdot (r-r^\prime)} \bigr)}
  {\tau + \mathbf{q}^2} \, . 
\end{equation}
The vertices contained in Eq.~(\ref{Hamilt}) are actually identical to those of
the DP field theory, whence the elements of the perturbation expansion coincide
with those depicted in FIG.~\ref{elemente} (with the changed propagator
(\ref{statProp}) and $\lambda = 1$). Correspondingly, determining the 
renormalization constants of the quasi-static theory proceeds as familiar from
DP, and is based on the same diagrams, see FIGS.~\ref{1-loop}, \ref{2-loop1},
and \ref{2-loop2}. Noticing that our conventions (\ref{Ren-Z}) imply that
$\mathring{\tilde{\varphi}} = \tilde{Z}^{1/2} \tilde{\varphi}$ and 
$\mathring{\varphi} = \tilde{Z}^{1/2}\varphi$, we explicitly find to two-loop 
order
\begin{subequations}
\label{Z-IP}
\begin{align}
  \tilde{Z} &= 1 + \frac{u}{6 \varepsilon} + \bigg( \frac{11}{\varepsilon^2} -
  \frac{37}{12 \varepsilon} \bigg) \frac{u^2}{36} + O(u^3) \, , \label{Z-IP1}\\
  Z_\tau &= 1 + \frac{u}{\varepsilon} + \bigg( \frac{9}{\varepsilon^2} -
  \frac{47}{12 \varepsilon} \bigg) \frac{u^2}{4} + O(u^3) \, , \label{Z-IP2}\\
  Z_u &= 1 + \frac{4 u}{\varepsilon} + \bigg(\frac{11}{\varepsilon^2} -
  \frac{59}{12 \varepsilon} \bigg) u^2 + O(u^3) \, . \label{Z-IP3}
\end{align}
These $Z$ factors coincide with those calculated in Ref.~\cite{Am76} for the 
Potts model in the single-state limit. The renormalization constants are known 
to three-loop order \cite{AKM81}. The RG functions appearing in the RGE read
to two-loop order \cite{Am76,Ja85}
\end{subequations}
\begin{subequations}
\label{RG-FudIP}
\begin{align}
  \tilde{\gamma}(u) &= - \frac{1}{6} \, u + \frac{37}{216} \, u^2 + O(u^3) \, ,
  \label{gammat-dIP}\\
  \kappa(u) &= \frac{5}{6} \, u - \frac{193}{108} \, u^2 + O(u^3) \, , 
  \label{kappa-dIP}\\
  \beta(u) & = \Big[ - \varepsilon + \frac{7}{2} \, u - \frac{671}{72} \, u^2
  + O(u^3) \Big] u \, . \label{beta-dIP}
\end{align}
\end{subequations}
The IR-stable fixed point, determined as zero of $\beta(u)$, is
\begin{equation}
\label{fixpt-dIP}
  u_\ast = \frac{2\varepsilon}{7} \Big[ 1 + \frac{671}{882} \, \varepsilon + 
  O(\varepsilon^2) \Big] \, .
\end{equation}
Thus we recover the well-known critical exponents for isotropic percolation:
\begin{subequations}
\label{dIP-Exp}
\begin{align}
  \eta_p &= \tilde{\eta} = \tilde{\gamma}(u_\ast) = - \frac{\varepsilon}{21}
  \Big[ 1 +\frac{206}{441} \, \varepsilon + O(\varepsilon^2) \Big] \, ,
  \label{eta-dIP}\\
  \nu & = \bigl[ 2 - \kappa(u_\ast) \bigr]^{-1} = \frac{1}{2} + 
  \frac{5\varepsilon}{84} \Big[ 1 + \frac{589}{2205} \, \varepsilon +
  O(\varepsilon^2) \Big] \, , \label{nu-dIP}\\
  \beta &= \nu \, \frac{d-2+\eta_p}{2} = 1 - \frac{\varepsilon}{6} \Big[ 1 +
  \frac{61}{1764} \, \varepsilon + O(\varepsilon^2) \Big] \, . \label{bet-dIP}
\end{align}
\end{subequations}

\subsubsection{Static observables}

Let $\mathcal{P}(S)dS$ be the measure for the probability that the cluster mass
of the debris generated by a seed at the origin is between $S$ and $S + dS$. We
obtain
\begin{equation}
\label{P}
  \mathcal{P}(S) = \left\langle \delta\left( \int \! d^dr \, 
  \varphi(\mathbf{r}) - S \right) \exp[\tilde{\varphi}(\mathbf{0})] 
  \right\rangle \, .
\end{equation}
For the probability density $\mathcal{P}(S)$ of large clusters with $S \gg 1$
we may use the expansion of the exponential to first order (higher orders 
asymptotically only lead to subleading corrections) and find
\begin{equation}
\label{P-as}
  \mathcal{P}_{\text{as}}(S) = \left\langle \delta\left( \int \! d^dr \,
  \varphi(\mathbf{r}) - S \right) \tilde{\varphi}(\mathbf{0}) \right\rangle\, .
\end{equation}

The \emph{percolation probability} $P_\infty$ is defined as the probability for
the existence of an infinite cluster generated from a single seed. Hence
\begin{align}
\label{Probinf1}
  P_\infty &= 1 - \lim_{c \to 0^+} \int_0^\infty \! dS \, \mathrm{e}^{-cS} P(S)
  \nonumber\\
  &= 1 - \lim_{c \to 0^+} \left\langle \exp\left[ \tilde{\varphi}(\mathbf{0}) -
  c \int \! d^dr \, \varphi(\mathbf{r}) \right] \right\rangle \, . 
\end{align}
Via expanding $\exp[\tilde{\varphi}(\mathbf{0})]$ we arrive at the asymptotic
form \cite{Ja03}
\begin{equation}
\label{PerkProb-inf}
  P_\infty \simeq - \lim_{c \to 0^+} \left\langle \tilde{\varphi}(\mathbf{0})
  \, \mathrm{e}^{- c \mathcal{M}} \right\rangle \, , 
\end{equation}
where $\mathcal{M} = \int \! d^dr \,\varphi(\mathbf{r})$. The virtue of this
formula is that it relates the percolation probability in an unambiguous manner
to an expression accessible by field theory. For actual calculations the term
$\exp(-cM)$ needs to be incorporated into the quasi-static Hamiltonian; i.e.,
we must replace the original $\mathcal{H}_{\rm IP}$ with
\begin{equation}
\label{newHamilt}
  \mathcal{H}_{{\rm IP},c} = \mathcal{H}_{{\rm IP}} + \int \! d^dr \, 
  c(\mathbf{r}) \varphi(\mathbf{r}) \, .
\end{equation}
Here, $c(\mathbf{x}) = c$ plays the role of a source conjugate to the field 
$\varphi$. Whereas in general $\langle \tilde{\varphi} \rangle = 0$ by 
causality if $c=0$, the limit $c \to 0^+$ leads to a non-vanishing order 
parameter $P_\infty$ in the active phase with spontaneously broken symmetry.
In terms of averages $\langle \cdots \rangle_c$ with respect to the new 
Hamiltonian (\ref{newHamilt}), we may now write 
\begin{equation}
\label{PerkProb1}
  P_\infty = -\lim_{c \to 0^+} \langle \tilde{\varphi}(\mathbf{0}) \rangle_c =
  - G_{0,1}(\mathbf{0};\tau, c \to 0^+, u,\mu) \, .
\end{equation}
With the aid of Eq.~(\ref{PerkProb1}) and the scaling form (\ref{As-Skal}) we 
readily obtain that
\begin{equation}
\label{Probinf}
  P_\infty \sim \theta(-\tau) |\tau|^\beta \, .
\end{equation}

In order to examine the scaling behavior of $\mathcal{P}(S)$, we consider its 
moments
\begin{equation}
\label{Debr-Mom}
  \langle S^k \rangle = \int_0^\infty \! dS \, S^k \, \mathcal{P}(S) \, .
\end{equation}
Using Eq.~(\ref{P-as}), our general scaling result (\ref{As-Skal}) implies
\begin{equation}
\label{Moments}
  \langle S^k \rangle \simeq \int (d^dr)^k \, G_{k,1}(\{\mathbf{r}\},
  \mathbf{0},\tau) \sim |\tau|^{\beta - k (d\nu-\beta)} \, ,
\end{equation}
which tells us that
\begin{equation}
\label{Pskal}
  \mathcal{P}_{\text{as}}(S,\tau) = S \, n_S(\tau) = 
  S^{1-\tau_p} \, f(\tau S^{\sigma_p}) \, , 
\end{equation}
where $n_S$ is the number of clusters of size $S$ per lattice site. These
\emph{cluster numbers} $n_S$ play an important role in percolation theory, 
which conventionally employs the scaling exponents $\tau_p$ and $\sigma_p$ as
defined in Eq.~(\ref{Pskal}). In terms of our earlier critical exponents, these
are
\begin{equation}
\label{Perk-Exp}
  \sigma_p = \frac{1}{d\nu - \beta} \, , \qquad
  \tau_p = 2 + \frac{{\beta}}{d\nu - \beta} \, .
\end{equation}
It follows from Eq.~(\ref{Moments}) that the mean cluster mass 
$\langle S \rangle$ (taken as average of the finite clusters) scales as
\begin{equation}
\label{MeanMass}
  \langle S \rangle = M(\tau) = M_0 \,|\tau|^{-\gamma} \, , \qquad 
  \gamma = d \nu - 2 \beta \, .
\end{equation}

Next we consider correlation functions restricted to clusters of given mass 
$S$.  In terms of the conventional unrestricted averages with respect to
$\mathcal{H}_{\rm IP}$, these restricted correlation functions can be expressed
for large $S$ as
\begin{equation}
\label{StatKorr}
  C_N^{(S)}(\{\mathbf{r}\},\tau) = \left\langle \varphi(\mathbf{r}_1) \ldots
  \varphi(\mathbf{r}_N) \ \delta\left( \int \! d^dr \, \varphi(\mathbf{r}) - S
  \right) \tilde{\varphi}(\mathbf{0}) \right\rangle^{(\text{conn})} \, .
\end{equation}
Their scaling from can again be read off from Eq.~(\ref{As-Skal}),
\begin{equation}
\label{KorrelFu}
  C_N^{(S)}(\{\mathbf{r}\},\tau) = |\tau|^{N \beta + d \nu} 
  F_N\big( \{|\tau|^\nu \mathbf{r}\},|\tau|^{d \nu - \beta} S \big) \, . 
\end{equation}
By means of these restricted Green functions we can write the \emph{radius of 
gyration} (i.e., the mean-square cluster radius) of clusters of size $S$ as
\begin{equation}
\label{R_s}
  R_S^2 = \frac{\int \! d^dr \, \mathbf{r}^2 \, C_1^{(S)}(\mathbf{r},\tau)}
  {2 d \int \! d^dr \, C_1^{(S)}(\mathbf{r},\tau)} \, ,
\end{equation}
whence with Eq.~(\ref{KorrelFu})
\begin{equation}
\label{GyrRad}
  R_S^2 = S^{2 / D_f} \, f_R(\tau \, S^{\sigma_p}) \, , 
\end{equation}
with the \emph{fractal dimension}
\begin{equation}
\label{FraktDim}
  D_f = d - \beta / \nu \, . 
\end{equation}

We conclude this section by considering the scaling behavior of the debris 
statistics if the initial state is prepared with a homogeneous seed density
$\rho_0 = h$. As discussed above, at the level of the quasistatic Hamiltonian 
$\mathcal{H}_{\rm IP}$ such an initial state translates to a further additive 
contribution $- h \int \! d^d r \, \tilde{\varphi}(\mathbf{r})$. Our general 
scaling form (\ref{As-Skal}) implies that the correlation functions of the 
densities $\varphi(\mathbf{r})$ for the case of a homogeneous initial condition
behave as
\begin{align}
\label{KorFuhom2}
  G_{N,\tilde{N}}(\{\mathbf{r}\},\tau,h) &= \sum_{k=0}^\infty 
  \frac{h^k}{k!} \int \! (d^d\widetilde{r})^k \, 
  G_{N,\widetilde{N}+k}(\{\mathbf{r}\},\{\widetilde{\mathbf{r}}\},\tau)
  \nonumber\\
  &= |\tau|^{\beta (N+\tilde{N})} F_N^\pm\big( \{|\tau|^\nu \mathbf{r}\},
  |\tau|^{\beta - d\nu} h \big) \, . 
\end{align}
It is obvious that the initial seed density $h$ plays the role of an ordering
field. It corresponds to the ghost field of conventional percolation theory \cite{StAh92BuHa96}. Hence, the Green functions do not display critical singularities as long
as $h > 0$. For a homogeneous initial condition the appropriate order parameter
is given by the debris density
\begin{equation}
\label{rho-stat}
  M = \langle \varphi(\mathbf{r}) \rangle_h = G_{1,0}(\mathbf{0},\tau,h) =
  |\tau|^\beta f_\rho^\pm\big( |\tau|^{\beta - d \nu} h \big) \, . 
\end{equation}
In the non-percolating phase ($\tau > 0$), the order parameter $M$ is linear in
$h$ for small seed density, with a susceptibility coefficient that diverges as
$\tau \to 0$,
\begin{equation}
\label{nDichte}
  M(\tau > 0,h) \sim \tau^{-\gamma} h \, , 
\end{equation}
with the susceptibility exponent $\gamma$ given in Eq.~(\ref{MeanMass}). At 
criticality ($\tau = 0$) the order parameter $M$ tends to zero for $h \to 0$ as
\begin{equation}
\label{cDichte}
  M(\tau = 0,h) \sim h^{1/\delta} \, , 
\end{equation}
with the exponent
\begin{equation}
\label{delta-stat}
  \delta = \frac{d \nu- \beta}{\beta} = 1 + \frac{\gamma}{\beta} \, . 
\end{equation}
Finally, in the percolating phase ($\tau < 0$) the order parameter becomes
independent of the initial seed density in the limit $h \to 0$ and tends to 
zero with $\tau$ according to
\begin{equation}
\label{pDichte}
  M(\tau < 0) \sim |\tau|^\beta \, , 
\end{equation}
which of course is just Eq.~(\ref{MeanMass}). Eqs.~(\ref{Probinf}) and 
(\ref{pDichte}) establish explicitly that the two distinct order parameters, 
namely the debris density $M$ and the percolation probability $P_\infty$, both
scale with the identical exponent $\beta$ near criticality.

\subsubsection{Crossovers}

At this stage we are in a position to discuss \emph{crossovers} between the two
fundamental percolation transitions \cite{FTS94,JaSt00}. We have established 
earlier that the introduction of the memory term $\propto \tilde{n} m n$ in the
response functional (\ref{J2}) changes scaling behavior from DP to dIP. Hence, 
memory is clearly a relevant perturbation (in the RG sense) in DP and the 
process described by $\mathcal{J}_{\rm DP}$ asymptotically crosses over to the 
process described by $\mathcal{J}_{\rm dIP}$, both in $d$ spatial and the 
single temporal dimension. We have also argued above that the static asymptotic
behavior of isotropic percolation in $d$ dimensions is described by the 
quasi-static Hamiltonian $\mathcal{H}_{\rm IP}$. If we now imagine a forest 
fire as a special realization of isotropic percolation and introduce a strong 
wind blowing in a preferred direction $\mathbf{e}$, such a directional 
disturbance may be described by an additional term 
$\propto \tilde{\varphi}(\mathbf{e} \cdot \nabla) \varphi$ in 
Eq.~(\ref{Hamilt}). Upon comparing with the longitudinal part 
$\tilde{\varphi}(\mathbf{e} \cdot \nabla)^2 \varphi$ of the Laplacian 
contribution, we see that this additional directional term is more relevant for
the IR behavior. That is, if we scale the transverse part of the Laplacian as 
$\nabla_\bot^2 = \nabla^2 - (\mathbf{e} \cdot \nabla)^2 \sim \mu^2$, the 
IR-scaling dimensions become $(\mathbf{e} \cdot \nabla) \sim \mu^2$ and 
$(\mathbf{e} \cdot \nabla)^2 \sim \mu^4$. As a consequence, the contribution
$\tilde{\varphi} (\mathbf{e} \cdot \nabla)^2 \varphi$ is asymptotically
irrelevant, and the quasistatic Hamiltonian becomes
\begin{equation}
\label{crossHamil}
  \mathcal{H}_{\rm IP}^\prime = \int \! d^dr \, \Big\{ \tilde{\varphi} \Big[
  c \, (\mathbf{e} \cdot \nabla) + \tau - \nabla_\bot^2 + \frac{g}{2} \bigl(
  \varphi - \tilde{\varphi} \bigr) \Big] \varphi - k \, \tilde{\varphi} \Big\}
  \, ,
\end{equation}
where $c$ is a new parameter that describes the directionality of the 
percolation in $d$ dimension. It is now straightforward to show by means of 
simple rescaling that by identifying the longitudinal spatial direction with 
`time' and the $(d-1)$ transverse subspace directions with a new space, 
$\mathcal{H}_{\rm IP}^{\prime}$ transforms into the response functional for DP,
namely $\mathcal{J}_{\rm DP}$, in $(d-1)$ spatial and $1$ temporal dimension. 
Thus, we have established the following schematic crossover scenarios:

$ \begin{array}
[c]{lllllll}
\ \text{DP} &  & \ \text{dIP} &  & \ \text{IP} &  & \ \text{DP} \\
\mathcal{J}_{\rm DP} & \quad\underrightarrow{\text{memory}} \quad & 
\mathcal{J}_{\rm dIP} & \quad\underrightarrow{
\begin{array}[c]{c} \text{quasistatic} \\ \text{limit} \end{array}}
\quad & \mathcal{H}_{\rm IP} & \quad\underrightarrow{\text{directionality}}
\quad & \mathcal{J}_{\rm DP} \\
d+1 &  & d+1 &  & \ d &  & (d-1)+1
\end{array} $

We note that one has to apply some care when trying to capture these crossover
scenarios by means of the dynamic RG, since both the transitions from DP to dIP
as well as from IP to DP involve different upper critical dimensions. Therefore
a mere $\varepsilon$ expansion about the $d_c$ of either theory cannot possibly
access the opposite scaling limit. One can, however, work out the RG flow 
functions that incorporate the entire crossover regimes at fixed dimension $d$,
provided one employs the correlation length $\xi$ as independent variable 
rather than $\tau$ in order to eliminate IR singularities. For more details on 
this method to describe the crossover from isotropic to directed percolation, 
see Refs.~\cite{FTS94,JaSt00}.

\subsection{Dynamic Observables}

In order to investigate scaling properties of genuinely dynamical observables 
of dIP, one must resort to the full response functional (\ref{JdIP}). For the 
determination of the last independent renormalization factor $Z$ to two-loop 
order, one needs to evaluate self-energy diagrams such as FIG.~\ref{1-loop}(a) 
and FIG.~\ref{2-loop1}, but now with one temporally delocalized leg of the 
vertex corresponding to the coupling $\propto \tilde{s} S s$ in 
$\mathcal{J}_{\rm dIP}$. We thus obtain from the renormalization of the 
derivative $\partial\Gamma_{1,1} / \partial \omega |_{\mathbf{q}=\omega=0}$ 
\cite{Ja85}
\begin{equation}
\label{dynZ-dIP}
  (Z \tilde{Z})^{1/2} = 1 + \frac{3 u}{4 \varepsilon} + \Big( 
  \frac{102}{\varepsilon} - \frac{227}{6} + 5 \ln 4 - 9 \ln3 \Big)
  \frac{u^2}{64 \varepsilon} + O(u^3) \, .
\end{equation}
In combination with Eqs.~(\ref{Z-IP1}) and (\ref{Ren-Z}), Eq.~(\ref{RG-Fu}) 
then yields the additional RG function
\begin{equation}
\label{zeta-dIP}
  \zeta(u) = - \frac{7u}{12} + \Big( \frac{1747}{54} + 9 \ln3 - 5 \ln4 \Big)
  \frac{u^2}{32} + O(u^3) \, . 
\end{equation}
Inserting the fixed point value (\ref{fixpt-dIP}) into Eq.~(\ref{FP-Werte}), we
find for the dynamic exponent
\begin{equation}
\label{z-dIP}
  z = 2 - \frac{\varepsilon}{6} - \Big[ \frac{937}{588} + \frac{9}{98} 
  (5 \ln 4 - 9 \ln 3) \Big] \frac{\varepsilon^2}{36} + O(\varepsilon^3) \, .
\end{equation}

The scaling form of the survival probability is equal to the expression 
(\ref{P(t)}), with the spreading exponent $\delta_s = \beta / \nu_{\Vert}$ and 
the longitudinal correlation exponent $\nu_{\Vert} = \nu z$ that follow from
the fundamental exponents listed in Eqs.~(\ref{dIP-Exp}) and (\ref{z-dIP}). 
Likewise, the radius of gyration is given by Eq.~(\ref{R(t)}) with the 
spreading exponent
\begin{equation}
\label{z-dIP-spr}
  z_s = \frac{2}{z} = 1 + \frac{\varepsilon}{12} + \Big[ \frac{1231}{294} +
  \frac{9}{49} (5 \ln 4 - 9 \ln3) \Big] \frac{\varepsilon^2}{144} + 
  O(\varepsilon^3) \, .
\end{equation}
Remarkably, this two-loop approximation for $z_s$ differs only by $1$-$2 \%$
from simulation results obtained in the physical dimensions $d=2$ and $d=3$ 
\cite{StAh92BuHa96,CaGra85}.

The number of active particles generated by a seed becomes
\begin{equation}
\label{N(t)-dIP}
  N(t,\tau)=t^{\theta_s - 1} f_N(\tau \, t^{1/\nu_{\Vert}}) \, , \qquad 
  \theta_s = (d \nu - 2 \beta) / \nu_{\Vert} \, .
\end{equation}
The active density $\rho(t,\tau,\rho_0)$ at time $t$, initialized by a finite 
homogeneous density $\rho_0$, is
\begin{equation}
\label{rho-dIP}
  \rho(t,\tau,h) = t^{-\delta_s - 1} 
  f_\rho(\tau \, t^{1/\nu_{\Vert}}, h \, t^{\theta_i + \delta_s+1}) \,  
\end{equation}
with an initial scaling exponent
\begin{equation}
\label{init-exp-dIP}
  \theta_i = (2-z-\eta_p) / z \, . 
\end{equation}

\section{Conclusions and Outlook: Other Classes of Percolation Processes}

In this overview, we have studied the field theory approach to percolating
systems. Based on the fundamental and universal features of the simple and
general epidemic processes, we have constructed a mesoscopic description in
terms of stochastic equations of motion, which we subsequently represented 
through a path integral with the dynamic response functional serving as the
appropriate effective action. We have also commented on a more microscopic
representation that starts from the classical master equation of a specific 
realization of such processes. In the bulk of this paper, we have provided a 
detailed description of the analysis of the ensuing stochastic field theories, 
from basic scaling properties to the perturbation expansion and UV 
renormalization, and explained why and how one may therefrom infer the correct 
asymptotic IR scaling behavior. We have derived a number of scaling relations 
for the critical exponents of directed (DP) and dynamic isotropic percolation 
(dIP), and explicitly demonstrated how these are linked to the large-scale, 
long-time properties of numerous static and dynamic observables. For the case 
of dIP, we have also derived the effective quasistatic field theory, which 
yields the scaling exponents of isotropic percolation. We remark that 
non-perturbative RG methods have recently been applied to the DP field theory 
as well \cite{CDDW04}.

Naturally, there are various possible extensions of the above models, some of 
which lead to novel critical properties. We end this review with a brief 
discussion of some interesting modifications of the standard percolation
processes.

\subsection{Long-Range Percolation}

In the standard version of the percolation processes in the language of an
infectious disease, the susceptible individuals can become contaminated by
already sick \emph{neighboring} individuals. At the same time sick individuals
are subject to spontaneous healing or immunization. In more realistic 
situations, however, the infection could be also long-ranged. As an example,
envision the spreading of a disease in an orchard where flying parasites 
contaminate the trees practically instantaneous in a widespread manner, 
provided the time scale of the parasites' flights is much shorter than the 
mesoscopic time scale of the epidemic process itself. Thus following a 
suggestion by Mollison \cite{Mol77}, Grassberger \cite{Gra86} introduced a 
variation of the epidemic processes with infection probability distributions 
$P(R) \propto 1 / R^{d+\sigma}$ which decay with the distance $R$ according to 
a power law. We will refer to such long-range distributions as \emph{L\'evy 
flights}, although a true L\'evy flight is defined by the Fourier transform,
$\tilde{P}(\mathbf{q}) \propto \exp( - b |\mathbf{q}|^\sigma)$ \cite{MoWe79}, 
and only L\'evy exponents in the interval $0<\sigma\leq2$ give rise to positive
distributions \cite{Bo59}. 

The spreading probability in this situation is rendered non-local,
\begin{equation}
\label{InfRate}
  \frac{\partial n(\mathbf{r},t)}{\partial t} \bigg\vert_{\rm spread} = 
  \int \! d^dr^\prime \, P(|\mathbf{r-r}^\prime|) \, n(\mathbf{r}^\prime,t)\, .
\end{equation}
After Fourier transformation of this equation, and after applying a small 
momentum expansion, we arrive at
\begin{equation}
\label{FouInfR}
  \frac{\partial n(\mathbf{q},t)}{\partial t} \bigg\vert_{\rm spread} = 
  \left[ p_0 - p_2 \, q^2 + p_\sigma \, q^\sigma + O\bigl( q^2,q^\sigma \bigr)
  \right] n(\mathbf{q},t) \, ,
\end{equation}
where the analytical terms stem from the short-range part of $P(R)$, whereas 
the non-analytical contributions arise from the power-law decay. The constant 
$p_0$ is included in the reaction rate as a negative (``birth'') contribution 
to $\tau$, while $p_2 \, q^2$ represents a diffusional term. In order to decide
which of the terms in Eq.~(\ref{FouInfR}) are relevant, one has to compare with
the scaling behavior of the Fourier-transformed susceptibility 
$\chi(\mathbf{q},\omega) \propto q^{2-\bar{\eta}}$, where $\bar{\eta}$ denotes 
the anomalous field dimension within the short-range spreading theories defined
by the response functionals (\ref{J3}), i.e., $\bar{\eta}$ is given by 
Eqs.~(\ref{eta-DP}) and Eq.~(\ref{eta-dIP}), respectively. If 
$\sigma < 2 - \bar{\eta}$, the parameter $p_\sigma$ constitutes a relevant 
perturbation and must be included in a renormalization group procedure. Upon 
taking the leading non-analytical term in Eq.~(\ref{FouInfR}) into account, the
harmonic part of the response functionals (\ref{J3}) changes to
\begin{subequations}
\begin{equation}
\label{Harm-LR}
  \mathcal{J}_0\left[ \tilde{s},s \right] = \int \! d^dr \, dt \, \Big\{
  \tilde{s} \Big( \partial_t + \lambda \Big[ \tau - \nabla^2 + f \bigl( 
  -\nabla^2 \bigr)^{\sigma/2} \Big] \Big) s \Big\} \, . 
\end{equation}
\end{subequations}

The mathematical treatment of this field theory with both gradient terms is 
somewhat delicate. Therefore, DP and dIP with long-range spreading were 
originally studied via Wilson's momentum shell RG method, since this approach
can handle relevant and irrelevant contributions on equal footing 
\cite{JOWH99}. Within a one-loop calculation, it was shown that the spreading 
is dominated by the L\'evy flights for $\sigma < 2 - \bar{\eta}$, and the
scaling exponents change continuously to their short-range counterparts at 
$\sigma = 2 - \bar{\eta}$. The application of renormalized field theory has to 
distinguish between two separate cases. If $2 - \sigma = O(\varepsilon)$, one 
must apply a double expansion in $\varepsilon$ and $\alpha = 2 - \sigma$,
following the work by Honkonen and Nalimov \cite{HoNa89}. Thereby the 
renormalizations were obtained to two-loop order \cite{Ja98-up}, and the 
one-loop results indeed corroborate the findings within the momentum shell 
approach with respect to the crossover to short-range percolation. 

In the case $2 - \sigma = O(1)$, the diffusional term becomes IR-irrelevant,
and must be removed in order to obtain a UV-renormalizable field theory in the 
infinite-cutoff limit. Then, by rescaling of time, $f$ can be set to $1$. The 
usual scaling $\mathbf{r} \sim \mu^{-1}$ yields $g^2 \sim \mu^\varepsilon$, and
$(\lambda t)^{-1} \sim \tau \sim \mu^\sigma$, where 
$\tilde{s} \sim s \sim \mu^{d/2}$ with $\varepsilon = 2 \sigma - d$ for DP and
$\tilde{s} \sim \mu^{(d-\sigma)/2}$, $s \sim \mu^{(d+\sigma)/2}$ with
$\varepsilon = 3 \sigma - d$ for dIP. Thus we infer $d_c = 2 \sigma$ and
$d_c = 3 \sigma$, respectively, to be the upper critical dimensions. The
propagator is now 
$G(\mathbf{q},t) = \theta(t) \exp[- \lambda (\tau +q^\sigma) t]$. The vertex
functions are analytical functions of the external momenta and frequencies as
long as $\tau > 0$. Thus, the non-analytic L\'evy flight term in
Eq.~(\ref{Harm-LR}) proportional to $(-\nabla^2)^{\sigma/2}$ does not require
renormalization, whence besides $\tilde{Z} = Z$ as usual we find for DP that
$Z_\lambda = 1$ exactly, while for long-range dIP $\tilde{Z} = Z_\lambda = 1$.
It turns out that counterterms are only needed for vertex functions with zero 
external momenta. The following identity
\begin{equation}
\label{Ident}
  \int \! d^dq \, f(q^\sigma) = \frac{2}{\sigma} \, \pi^{(d-d^{\prime})/2} \,
  \frac{\Gamma(d/\sigma)}{\Gamma(d/2)} \int \! d^{d^{\prime}}k \, f(k^2) \, ,
\end{equation}
where $d^{\prime} = 2 d / \sigma$, is useful for the explicit computation of 
the $Z$ factors, from which subsequently the RG functions (\ref{RG-Fu}) are
found. To one-loop order for long-range DP, the result is
\begin{subequations}
\label{RG-FulrDP}
\begin{align}
  \beta(u) &= \Bigl[ - \varepsilon + \frac{7u}{4} + O(u^2) \Bigr] u \, ,
  \label{RG-FulrDP1} \\
  \kappa(u) &= \frac{u}{2} + O(u^2) \, , \quad 
  \tilde{\gamma}(u) = \gamma(u) = \zeta(u) = - \frac{u}{4} + O(u^2) \, ,
  \label{RG-FulrDP2}
\end{align}
\end{subequations}
whereas for long-range dIP
\begin{subequations}
\label{RG-FulrdIP}
\begin{align}
  \beta(u) &= \bigl[ - \varepsilon + 4u + O(u^2) \bigr] u \, , 
  \label{RG-FulrdIP1}\\
  \kappa(u) &= u + O(u^2) \, , \quad \tilde{\gamma}(u) = 0 \, , \quad 
  \gamma(u) = 2 \zeta(u) = - \frac{3u}{2} + O(u^2) \, . \label{RG-FulrdIP2}
\end{align}
\end{subequations}
At the long-range DP and dIP fixed points 
$u_{\ast} = 4 \varepsilon / 7 + O(\varepsilon^2)$ and 
$u_{\ast} = \varepsilon / 4 + O(\varepsilon^2)$, respectively, we then get the 
critical exponents
\begin{subequations}
\label{lr-Exp}
\begin{align}
  \text{long-range DP:}\quad &\tilde{\eta} = \eta = z - \sigma
  = - \frac{\varepsilon}{7} + O(\varepsilon^2) \, , \quad
  \frac{1}{\nu} = \sigma - \frac{2 \varepsilon}{7} + O(\varepsilon^2) \, ,
  \label{lr-Exp-DP1}\\
  &\beta = \nu \, \frac{d + \eta}{2} = 1 - \frac{2 \varepsilon}{7 \sigma} +
  O(\varepsilon^2) \, , \label{lr-Exp-DP2}\\
  \text{long-range dIP:}\quad &\eta = 2 (z - \sigma) 
  = -\frac{3 \varepsilon}{8} + O(\varepsilon^2) \, ,\quad
  \frac{1}{\nu} = \sigma - \frac{\varepsilon}{4} + O(\varepsilon^2) \, ,
  \label{lr-Exp-dIP1}\\
  &\beta = \nu \, \frac{d-\sigma}{2} = 1 - \frac{\varepsilon}{4 \sigma}
  +O(\varepsilon^2) \, , \quad \tilde{\eta} = 0 \, . \label{lr-Exp-dIP2}
\end{align}
\end{subequations}

\subsection{Percolation Boundary Critical Behavior}

Within the field theory formulation, one can also investigate the influence of
a spatial boundary on critical behavior \cite{Di86,Di97}. Generally, for
percolation processes four possible scenarios can be envisioned: The boundary
remains inactive, whereas the bulk is critical (\emph{ordinary} transition),
the boundary is active, the bulk is critical (\emph{extraordinary} transition),
the boundary is critical, but the bulk inactive (\emph{surface} transition), or
both boundary and bulk are critical, which represents a multicritical point
(the \emph{special} transition) \cite{JSS88b,JSS88c,FHL01}. Let us consider a 
semi-infinite geometry, with a boundary plane at $z = 0$; in this situation we
need to supplement the dynamic response functionals (\ref{J3}) with the surface
action
\begin{equation}
\label{surface}
  \mathcal{J}_{\rm surf} = \int \! d^{d-1}r \, dt \, \lambda \, \tau_s \, 
  \tilde{s}(z=0) \, s(z=0) \, ,
\end{equation}
and impose the boundary condition $\partial_z s |_{z=0} = \tau_s \, s(z=0)$.
Naive power counting yields that the new parameter $\tau_s$ is relevant, 
whence its RG fixed points are either $0$ or $\pm \infty$. The ordinary 
transition scenario corresponds to $\tau_s \to +\infty$ and the special 
transition to $\tau_s \to 0$. The presence of the boundary implies a different 
scaling behavior of the fields near the surface as compared to the bulk. For 
example, the surface order parameter acquires new critical exponents
\begin{subequations}
\label{surfexp}
\begin{align}
  \text{DP}: \quad \beta^{(\rm o)}_1 &= \frac{3}{2} - \frac{7 \varepsilon}{48} 
  + O(\varepsilon^2) \, , \quad  
  \beta^{(\rm s)}_1 = 1 - \frac{\varepsilon}{4} +O(\varepsilon^2)\, ,\\
  \text{dIP}: \quad \beta^{(\rm o)}_1 &= \frac{3}{2} - \frac{11\varepsilon}{84}
  + O(\varepsilon^2) \, , \quad  
  \beta^{(\rm s)}_1 = 1 - \frac{3\varepsilon}{14} + O(\varepsilon^2) \, .
\end{align}
\end{subequations}

\subsection{Multispecies Directed Percolation Processes}

One may readily generalize the previous mesoscopic description of the simple
epidemic process to multiple activity carriers $\alpha=1,2,\ldots$ in order to 
capture, say, a variety of population species near an extinction threshold. In 
the spirit of Sec.~2.1 one thus arrives at the coupled DP Langevin equations 
\cite{Ja01}
\begin{subequations}
\label{MultLange}
\begin{align}
  \partial_t \, n_\alpha &= \lambda_\alpha \, \nabla^2 n_\alpha 
  + R_\alpha[\{ n_\alpha \}] + \zeta_\alpha \, , \label{multL1}\\
  R_\alpha[\{ n_\alpha \}] &= - \lambda_\alpha \, \Bigl( \tau_\alpha + 
  \frac{1}{2} \sum_\beta g_{\alpha \beta} n_\beta + \ldots \Bigr) \, , 
  \label{Multdens}
\end{align}
\end{subequations}
with the stochastic noise correlations
\begin{equation}
\label{Multnoise}
  \overline{\zeta_\alpha(t) \, \zeta_\beta(t^\prime)} = \lambda_\alpha \,
  g_\alpha \, \delta_{\alpha \beta} \, n_\alpha(t) \, \delta(t-t^\prime) \, .
\end{equation}

Although this coupled multispecies systems appears to be very rich, it turns 
out that in fact all the ensuing renormalizations are given precisely by those
of the single-species process, and hence the critical behavior at the 
extinction threshold is quite remarkably just that of DP again \cite{Ja01}. In
addition, this generically universal model displays an instability that 
asymptotically leads to unidirectionality in the interspecies couplings. A
special situation arises when several control parameters $\tau_\alpha$ vanish 
simultaneously, which implies \emph{multicritical} behavior of unidirectionally
coupled DP processes \cite{THH98}. In the active phase, one then finds a 
hierarchy of order parameter exponents $\beta_\alpha$ with 
\begin{equation}
\label{multiDP}
  \beta_1 = \beta_{\rm DP} = 1 - \frac{\varepsilon}{6} + O(\varepsilon^2) \, , 
  \quad \beta_2 = \frac{1}{2} - \frac{13\varepsilon}{96} + O(\varepsilon^2) \, ,
  \ldots , \quad \beta_k = \frac{1}{2^k} - O(\varepsilon) \, . 
\end{equation}
In addition, one can show that the crossover exponent associated with the 
multicritical point is $\Phi \equiv 1$ to all orders in perturbation theory 
\cite{Ja01}. There remains, however, an unresolved technical issue originating 
from the emergence of a relevant coupling that enters the perturbation 
expansion \cite{THH98}. Quite analogous features also characterize multispecies
generalizations of the general epidemic process or coupled dIP processes.

\subsection{Directed Percolation with a Diffusive Conserved Field}

Several years ago Kree, Schaub, and Schmittmann \cite{KSS89} introduced a model
that consists of the two-species reaction-diffusion system $B \to 2B$, 
$2B \to B$, and $B+C \to C$ with unequal diffusion constants species $B$ and 
$C$. The active particles $B$, whose density we set proportional to 
$s(\mathbf{r},t)$, become poisoned by a diffusing conserved quantity $C$ with
density $c(\mathbf{r},t)$. In a mesoscopic description, this KSS-model is 
described by the coupled Langevin equations
\begin{subequations}
\label{Beweg}
\begin{align}
  \lambda^{-1} \partial_t s &= \nabla^2 s - \Big( \tau + \frac{g}{2} \, s
  + f \, c \Big) s + \zeta_s \, , \label{Beweg_s}\\
  \gamma^{-1} \partial_t c &= \nabla^2 c + \zeta_c \, , \label{Beweg_c}
\end{align}
\end{subequations}
with positive parameters $\lambda$, $\gamma$, and $g$. The stochastic forces
$\zeta_i(\mathbf{r},t)$ must respect the absorbing state condition as well as
the conservation property. Hence their Gaussian fluctuations are given by
\begin{subequations}
\label{Lang}
\begin{align}
  \overline{\zeta_s(\mathbf{r},t) \, \zeta_s(\mathbf{r}^\prime,t^\prime)} &=
  \lambda^{-1} \tilde{g} \, s(\mathbf{r},t) \, 
  \delta(\mathbf{r}-\mathbf{r}^\prime) \, \delta(t-t^{\prime}) \, ,
  \label{Lang_s}\\
  \overline{\zeta_c(\mathbf{r},t) \, \zeta_c(\mathbf{r}^\prime,t^\prime)} &=
  \tilde{\gamma}^{-1}(-\nabla^2) \, \delta(\mathbf{r}-\mathbf{r}^\prime) \,
  \delta(t-t^\prime) \, ,\label{Lang_c}\\
  \overline{\zeta_s(\mathbf{r},t) \, \zeta_c(\mathbf{r}^\prime,t^\prime)} &= 0
  \, . \label{Lang_sc}
\end{align}
\end{subequations}
More recently, van Wijland, Oerding, and Hilhorst studied the two-species
reaction-diffusion system $A+B \to 2B$ and $B \to A$ with unequal 
diffusivities \cite{WOH98}. Here, the total number of $A$ and $B$ particles 
constitutes a conserved quantity $C$. Elimination of the $A$ density in favor 
of the density of $C$ then recovers the stochastic equations of motion 
(\ref{Beweg}), (\ref{Lang}), yet now with a cross-diffusion term in 
Eq.~(\ref{Beweg_c}):
\begin{equation}
\label{cross-diff}
  \gamma^{-1} \partial_t c = \nabla^2 (c-\sigma s) + \zeta_c \, , 
\end{equation}
where $\sigma$ is proportional to the difference of the diffusion constants of
the $B$ and $A$ species.

Both the KSS and the WOH model represent generalizations of DP via including
the influence of a non-critical conserved quantity in the critical dynamics of 
the agent, akin to the generalization of the relaxational model A of 
near-equilibrium critical dynamics to model C \cite{HH77}. Therefore we propose
the label DP-C for this modification of the DP universality class. Following
the same lines that resulted in the response functional (\ref{J2}), and 
omitting IR-irrelevant terms, we obtain the renormalizable response functional
of the DP-C class corresponding to the above Langevin equations,
\begin{align}
\label{JDP-C}
  \mathcal{J} &= \int \! d^dr \, dt \, \biggl\{ \tilde{s} \Bigl[ \partial_t +
  \lambda (\tau - \nabla^2 + f c) + \frac{\lambda}{2} \, 
  (g s -\tilde{g} \tilde{s}) \Bigr] s \nonumber\\
  &\qquad\qquad\quad + \tilde{c} \Bigl[ \partial_t c - \gamma \nabla^2 
  (c - \sigma s) \Bigr] - \tilde{\gamma} \, (\nabla \tilde{c})^2 \biggr\} \, .
\end{align}

Our first observation concerns the stability of DP-C. Consider the mean-field
approximation for the stationary state. Eq.~(\ref{cross-diff}) implies that 
$c(\mathbf{r}) = \sigma s(\mathbf{r}) + c_0$. Including the constant $c_0$ in 
the variable $\tau$, we arrive at
\begin{equation}
\label{g-eff}
  \nabla^2 s - \Big( \tau + \frac{g_{\rm eff}}{2} \, s \Big) s = 0 \, , \qquad 
  g_{\rm eff} = g + 2 \sigma f \, ,
\end{equation}
which demonstrates that the stable homogeneous solution $s = 0$ for positive
$\tau$ and $g_{\rm eff}$ becomes unstable for $g_{\rm eff} \leq 0$. Higher 
orders of $s$ should then be included in the density expansion, and one
would expect the transition to become dicontinuous or first-order. A continuous
second-order transition therefore requires that the constraint 
$g > -2 \sigma f$ be satisfied. An other qualitative view on this instability 
is illustrated by the following consideration. Assume $f \geq 0$ in the 
following, i.e., the density $c$ operates as an inhibitor. If now an 
enhancement of $s$ is created by a fluctuation in some region of space the 
current contribution $\mathbf{j}_{\rm cross} = \gamma \sigma \nabla s$ shows 
that the inhibitor flows into this region if $\sigma > 0$ and subsequently 
reduces the fluctuation. However, if $\sigma < 0$, the inhibitor is reduced by 
the flow out of this region, and the fluctuation of $s$ becomes increasingly 
enhanced.

Rescaling the fields $c$, $\tilde{c}$, and the parameter $\sigma$, we may set 
$\tilde{\gamma} = \gamma$. The response functional (\ref{JDP-C}) possesses the 
following symmetries under three transformations that involve constant 
continuous parameters $\alpha_i$:
\begin{subequations}
\label{Symm-DPC}
\begin{align}
  \text{I:} \quad &\tilde{c} \to \tilde{c} + \alpha_1 \, ; \label{I}\\
  \text{II:} \quad &c \to c+\alpha_2 \, , \quad \tau \to \tau - f \alpha_2 \, ;
  \label{II}\\
  \text{III:} \quad &s \to \alpha_3 \, s \, , \quad \tilde{s} \to \alpha_3^{-1}
  \tilde{s} \, , \quad \sigma \to \alpha_3^{-1} \sigma \, , \quad 
  g \to \alpha_3^{-1} g \, , \quad \tilde{g} \to \alpha_3 \, \tilde{g} \, .
  \label{III}
\end{align}
\end{subequations}
Moreover, $\mathcal{J}$ is invariant under the inversion
\begin{equation}
\label{IV}
  \text{IV:} \quad \tilde{c} \to - \tilde{c} \, , \quad c \to - c \, , \quad
  \sigma \to - \sigma \, , \quad f \to - f \, . 
\end{equation}
In the particular case $\sigma = 0$, the time reflection
\begin{align}
  \text{V:} \quad &\sqrt{g/\tilde{g}} \, s(\mathbf{r},t) \leftrightarrow 
  - \sqrt{\tilde{g}/g} \, \tilde{s}(\mathbf{r},-t) \, , \label{Va}\\
  &c(\mathbf{r},t) \to c(\mathbf{r},-t) \, , \quad \tilde{c}(\mathbf{r},t)
  \to c(\mathbf{r},-t) - \tilde{c}(\mathbf{r},-t) \label{Vb}
\end{align}
yields an additional discrete symmetry transformation. The invariance with
respect to the symmetry V distinguishes the special KSS model from the general 
DP-C class. 

Symmetry I results from the conservation property of the field $c$. Symmetries
III and IV show that dimensionless invariant coupling constants and parameters
are defined by $u = \tilde{g} g \, \mu^{-\varepsilon}$, 
$v = f^2 \mu^{-\varepsilon}$, $w = \sigma \tilde{g} f \mu^{-\varepsilon}$, and 
the ratio of the kinetic coefficients $\rho = \gamma / \lambda$ with 
$\varepsilon = 4-d$. Symmetry III implies that the response functional 
$\mathcal{J}$ again contains a redundant parameter which can be fixed in 
different ways. Dimensional analysis and the scaling symmetry III applied to 
the Green functions $G_{N,\tilde{N};M,\tilde{M}} = \langle [s]^N 
[\tilde{s}]^{\tilde{N}} [c]^M [\tilde {c}]^{\tilde{M}} \rangle$ gives
\begin{subequations}
\begin{align}
\label{Skalg}
  G_{N,\tilde{N};M,\tilde{M}} &= \alpha_3^{\tilde{N}-N} 
  G_{N,\tilde{N};M,\tilde{M}}(\{\mathbf{r},t\},\tau,\alpha_3^{-1} \sigma,
  \alpha_3 \, \tilde{g},\alpha_3^{-1} g,f,\lambda,\gamma,\mu)\\
  &= \sigma^{\tilde{N}-N} F_{N,\tilde{N};M,\tilde{M}}(\{\mu � \mathbf{r},
  \gamma \mu^2 t\},\mu^{-2} \tau,u,v,w,\rho) \label{SkalSig}\\
  &= \bigl( g/\tilde{g} \bigr)^{(\tilde{N}-N)/2} 
  F_{N,\tilde{N};M,\tilde{M}}^{\prime}(\{\mu \, \mathbf{r},\gamma \mu^2 t\},
  \mu^{-2} \tau,u,v,w,\rho) \, ,
\end{align}
\end{subequations}
wherein the UV singularities and critical properties reside in the functions
$F_{N,\tilde{N};M,\tilde{M}}$ and $F_{N,\tilde{N};M,\tilde{M}}^{\prime}$, 
respectively. If $\sigma = 0$, only Eq.~(\ref{Skalg}) can be used, and it is 
natural to apply a rescaling that leads to $g = \tilde{g}$, which is fixed also
under renormalization by the time inversion symmetry V. However, this is not
the case in the general situation $\sigma \neq 0$, whence it is more natural to
absorb $\sigma$ into the fields $s$ and $\tilde{s}$. In other words, one may 
then hold $\sigma = \mathring{\sigma}$ constant under renormalization.

It is easily seen that loop diagrams do not contribute to the vertex functions 
$\Gamma_{\tilde{N},N;\tilde{M},M}$ with $\tilde{M} \geq 1$, whence we infer
$\mathring{\tilde{c}} = \tilde{c}$, $\mathring{c} = c$, 
$\mathring{\gamma} = \gamma$, and $\mathring{\sigma} \mathring{s} = \sigma s$.
We define the remaining renormalizations via
\begin{subequations}
\label{Ren-DPC}
\begin{align}
  &\mathring{s} = Z^{1/2} s \, , \quad \mathring{\tilde{s}} = \tilde{Z}^{1/2}
  \tilde{s} \, , \quad \mathring{\lambda} = Z^{-1/2} \tilde{Z}^{-1/2} 
  Z_\lambda \lambda \, , \quad \mathring{\tau} = Z_\lambda^{-1} Z_{\tau} \tau 
  + \mathring{\tau}_c \, , \\
  &G_\varepsilon \mathring{\tilde{g}} \mathring{g} = Z^{-1/2} \tilde{Z}^{-1/2}
  Z_{\lambda}^{-2} \bigl( Z_u u + Y w \bigr) \mu^\varepsilon \, , \quad
  G_\varepsilon \mathring{f}^2 = Z_\lambda^{-2} Z_v v \mu^\varepsilon \, , \\
  &G_\varepsilon \mathring{\sigma} \mathring{\tilde{g}} \mathring{f} = Z^{-1/2}
  \tilde{Z}^{-1/2} Z_\lambda^{-2} Z_w w \mu^\varepsilon \, , \quad 
  Z_w^2 = Z_u Z_v \, ,
\end{align}
\end{subequations}
where the last setting can be implemented through an appropriate choice for 
$Y$. For $\sigma \neq 0$, the non-renormalization of $s$ ($Z=1$) follows from
$\sigma = \mathring{\sigma}$. If $\sigma = 0$, we have $Z = \tilde{Z}$. 
Symmetry II in connection with the trivial renormalization of $c$ shows that 
$f$ is renormalized with the same $Z$ factor as $\tau$: $Z_v = Z_\tau^2$.

The RGE (\ref{RGE}) now contains four Gell-Mann--Low functions $\beta_i$
corresponding to the four dimensionless parameters $u$, $v$, $w$, and $\rho$.
A one-loop calculation gives
\begin{subequations}
\label{beta-DPC}
\begin{align}
  \beta_u &= \Big[ - \varepsilon + \frac{3u}{2} 
  - \frac{2 (5 + 5 \rho + 2 \rho^2) v}{(1 + \rho)^3} 
  + \frac{(7 + 8 \rho + 3 \rho^2) w}{(1 + \rho)^3} + O(\text{2-loop}) \Big] u
  \nonumber\\
  &\qquad + \Big[ \frac{u}{1 + \rho} - \frac{4v}{\rho (1 + \rho)} 
  + \frac{2w}{\rho (1 + \rho)} + O(\text{2-loop}) \Big] w \, , 
  \label{beta-DPC-u}\\
  \beta_v &= (- \varepsilon + 2 \kappa) v \nonumber\\
  &= \Big[ - \varepsilon + \frac{3u}{4} - \frac{4v}{(1 + \rho)^3}
  + \frac{(9 + 8 \rho + 3 \rho^2) w}{2 (1 + \rho)^3} + O(\text{2-loop}) \Big]
  v \, , \label{beta-DPC-v}\\
  \beta_w &= \Big[ - \varepsilon + u 
  - \frac{2 (3 + 2 \rho + \rho^2) v}{(1 + \rho)^3}
  + \frac{(5 + 5 \rho + 2 \rho^2) w}{(1 + \rho)^3} + O(\text{2-loop}) \Big] w
  \, , \label{beta-DPC-w}\\
  \beta_\rho &= - \zeta \rho = \Big[ \frac{u}{8} - \frac{2v}{(1 + \rho)^3}
  + \frac{(7 + 4 \rho + \rho^2) w}{4 (1 + \rho)^3} + O(\text{2-loop}) \Big]
  \rho \, . \label{beta-DPC-rho}
\end{align}
\end{subequations}
The RG flow functions leading to the anomalous dimensions of the fields $s$ and
$\tilde{s}$ are found to be
\begin{subequations}
\label{gam-DPC}
\begin{align}
  \gamma &= \left\{ \begin{array}[c]{lll} \tilde{\gamma} \quad & \text{if} & \
  \sigma = 0 \\ 0 & \text{if} & \ \sigma \neq 0 \end{array} \right. \, ,
  \label{gam-DPC1}\\
  \gamma + \tilde{\gamma} &= \Big[ - \frac{u}{2} + \frac{4v}{(1 + \rho)^2}
  - \frac{(3 +\rho) w}{(1 + \rho)^2} \Big] + O(\text{2-loop}) \, . 
  \label{gam-DPC2}
\end{align}
\end{subequations}

At a non-trivial IR-stable fixed point with all $\beta_i = 0$ and $v_{\ast}$ 
and $\rho_{\ast}$ both different from $0$ and $\infty$, one finds from
Eqs.~(\ref{beta-DPC-v}) and (\ref{beta-DPC-rho}) the exact statements
$\kappa_{\ast} = \varepsilon/2$ and $\zeta_{\ast} = 0$, and Eq.~(\ref{As-Skal})
yields the asymptotic scaling properties of the Green's functions,
\begin{equation}
\label{GSkal}
  G_{N,\tilde{N};M,\tilde{M}}(\{\mathbf{r},t\},\tau) = l^{\delta_G} \,
  G_{N,\tilde{N};M,\tilde{M}}(\{l\mathbf{r},l^2 t\},\tau / l^{d/2}) \, .
\end{equation}
Consequently, the DP-C universality is characterized by the exact critical
exponents
\begin{equation}
  z = 2 \, , \qquad \nu = 2/d \, .
\end{equation}
In Eq.~(\ref{GSkal}) we have
\begin{subequations}
\label{beta_exp-DPC}
\begin{align}
  \delta_G &= \bigl[ (M + \tilde{M}) + (N \beta + \tilde{N} \beta^{\prime})
  \bigr] d/2 \, , \\
  \beta &= \left\{ \begin{array}[c]{lll} \beta^{\prime} \quad & \text{if} & \
  \sigma = 0 \\ 1 & \text{if} & \ \sigma \neq 0 \end{array} \right. \, , \\
  \beta^{\prime} &= (d+\eta^{\prime}) / d \, , \qquad 
  \eta^{\prime} = \tilde{\gamma}_{\ast} \, .
\end{align}
\end{subequations}
Thus, there is merely one unknown independent critical exponent $\eta^{\prime}$
in the DP-C universality class.

To order $\varepsilon$, one finds the following IR fixed points as zeros of the
Gell-Mann--Low functions $\beta_i$:
\begin{subequations}
\label{fix-DPC}
\begin{align}
  &\sigma=0: \quad \rho_{\ast} = \frac{1}{2} \, , \quad 
  u_{\ast} = 2 \varepsilon \, , \quad v_{\ast} = \frac{27}{64} \, \varepsilon
  \, , \quad w_{\ast} = 0 \, ; \\
  &\sigma > 0: \quad \rho_{\ast} = \bigl( 2 + \sqrt{3} \bigr)^{1/3} +
  \bigl( 2 - \sqrt{3} \bigr)^{1/3} - 2 \, , \\
  &\qquad\qquad\ u_{\ast} = \frac{4}{2 + \rho_{\ast}} \, \varepsilon \, , \quad
  v_{\ast} = \frac{1 + \rho_{\ast}}{4} \, \varepsilon \, , \quad 
  w_{\ast} = \frac{1 - 5 \rho_{\ast}}{\rho_{\ast}} \, \varepsilon \, ; \\
  &\sigma < 0: \quad \text{run-away flow.} 
\end{align}
\end{subequations}
The run-away solution in the case $\sigma < 0$ eventually violates the 
stability bound $u > 2w$ in the course of the flow, which indicates the
emergence of a  first-order transition is this situation. In the other cases,
we find for the remaining exponents to one-loop order:
\begin{subequations}
\label{eta-DPC}
\begin{align}
  &\sigma = 0: \quad \eta^{\prime} = - \varepsilon / 8 \, , \\
  &\sigma > 0: \quad \eta^{\prime} = - \frac{\varepsilon}{3 + \rho_{\ast}} \, .
\end{align}
\end{subequations}
All scaling laws now follow from the scaling of the Green's functions 
(\ref{GSkal}). As important examples we note the initial slip exponent 
$\theta_i = - \eta^{\prime} / 2$ and $\theta_i = -\eta^{\prime} / 4$ for 
$\sigma=0$ and $\sigma>0$, respectively.

\subsection{Directed Percolation and Quenched Disorder}

According to the equation of motion (\ref{Beweg_s}), the DP-C processes can be
understood as models of systems where the critical control parameter $\tau$
itself becomes a dynamical variable $\tau+fc(\mathbf{r},t)$ subject to a
diffusive process. For the case that a DP process evolves in a 
\emph{disordered medium}, the field $c$ represents a static but spatially 
inhomogeneous random quantity with short-range Gaussian correlations
$\overline{c(\mathbf{r}) \, c(\mathbf{r}^{\prime})} \propto 
\delta(\mathbf{r} - \mathbf{r}^{\prime})$. It is easily seen that the influence
of the disorder on the other parameters of the model and beyond-Gaussian 
correlations are IR-irrelevant. In contrast to equilibrium systems, where the 
normalization factor of the probability distribution, namely the partition 
function, is itself a functional of the disorder, the description of dynamics 
by means of an exponential weight $\exp(-\mathcal{J}_c)$ with a response 
functional $\mathcal{J}_c$ that depends on the disorder field $c(\mathbf{r})$ 
does not require a normalization factor. Hence, averaging over $c$ can be 
easily performed directly on the statistical weight directly, 
$\overline{\exp(-\mathcal{J}_c)} =: \exp(-\mathcal{J})$, without invoking any 
additional procedures such as, e.g., replication. One thereby obtains the 
following effective action for the evaluation of path integrals averaged over 
the randomness \cite{Ja97}:
\begin{equation}
\label{J-rand}
  \mathcal{J} = \int \! d^dr \, \Big\{ \int \! dt \, \tilde{s} \Big[ 
  \partial_t + \lambda \bigl( \tau-\nabla^2 \bigr) + \frac{\lambda g}{2} 
  \bigl( s - \tilde{s} \bigr) \Big] s - \frac{\lambda^2}{2} \, f 
  \Big[ \int \! dt \, \tilde{s} s \Big]^2 \Big\} \, . 
\end{equation}

From here, the calculation of the renormalizations and the RG flow functions 
proceeds in the usual manner. To one-loop order, one finds the Gell-Mann--Low
functions
\begin{equation}
\label{beta-rand}
  \beta_u = \Bigl( - \varepsilon + \frac{3 u}{2} - 10 v \Bigr) u \, , \qquad
  \beta_v = \Bigl( - \varepsilon + \frac{3 u}{4} - 8 v \Bigr) v \, , 
\end{equation}
where $v$ is the renormalized dimensionless coupling corresponding to $f$. It 
is now easily seen that the flow equations (\ref{beta-rand}) allow only for 
\emph{runaway} solutions as $l \to 0$ in the physical region $u>0$, $v>0$. The 
fixed point of the pure system $u_{\ast} = 2 \varepsilon / 3$, $v_{\ast}=0$ is 
unstable, as is the non-physical fixed point $u_{\ast} = - 4 \varepsilon / 9$,
$v_\ast = -1/6$. There is a stable fixed point, namely $u_{\ast}= 0$, 
$v_\ast = -1/8$, but it is in the non-physical region as well. The runaway 
RG trajectories render the perturbation expansion useless and, perhaps, 
indicate a more complicated critical behavior than just simple power laws. 
Indeed, the simulations by Moreira and Dickman \cite{MoDi96} show logarithmic
critical spreading instead of power laws, and seem to display violation of
simple scaling \cite{MoDi98}. Apparently the continuum field-theoretic 
treatment is not capable of capturing the strong disorder limit of DP, which 
may be dominated by rare localized excitations. A promising alternative 
approach has recently been developed that employs a real-space RG framework 
specifically tailored to strongly disordered systems \cite{HIV03}.

\section*{Acknowledgements}

We have benefitted from fruitful collaborations and insightful discussions with
numerous colleagues. Specifically, we would like to thank J.~Cardy, 
O.~Deloubri\`ere, H.W.~Diehl, V.~Dohm, G.~Foltin, E.~Frey, Y.~Goldschmidt, 
P.~Grassberger, H.~Hilhorst, H.~Hinrichsen, M.~Howard, R.~Kree, \"U.~Kutbay, 
S.~L\"ubeck, K.~Oer\-ding, B.~Schaub, B.~Schmittmann, F.~Schwabl, O.~Stenull, 
B.~Vollmayr-Lee, F.~van~Wijland, and R.~Zia for their valuable contributions.
U.C.T. gratefully acknowledges support from the National Science Foundation 
under Grant No.~DMR-0308548.

\end{document}